\documentclass[12pt]{iopart}

\usepackage{graphicx}
\usepackage{iopams}  
\usepackage{epsfig}
\usepackage{color}
\usepackage{hyperref}
\usepackage[normalem]{ulem}
\begin{document}

\title[]{Real-space effects of a quench in the Su-Schrieffer-Heeger model and elusive dynamical appearance of the topological edge states}

\author{Lorenzo Rossi, Fausto Rossi \& Fabrizio Dolcini }

\address{DISAT, Politecnico di Torino, corso Duca degli Abruzzi 24, 10129 Torino (Italy)}
\ead{lorenzo.rossi@polito.it}
\vspace{10pt}

\begin{abstract}
The topological phase  of the Su-Schrieffer-Heeger (SSH) model  is known to    exhibit two edge states  that are  topologically protected by the chiral symmetry.   We demonstrate that, for any parameter quench performed on the half-filled SSH chain, the occupancy of each lattice site remains  locked to $1/2$ at any time,  due to the additional time-reversal and charge conjugation symmetries.  In particular, for a quench from the trivial to the topological phase, no signature of  the topological edge states appears in real-space {occupancies}, independently of the    quench protocol, the  temperature of the pre-quench thermal state or the presence of chiral disorder.  
However,   a suitably designed local quench from/to a SSH ring threaded by a magnetic flux can break these additional symmetries while preserving the chiral one. Then, real-space effects of the quench do appear and exhibit   different dynamical features in the topological and in the trivial phases. Moreover,  when  the particle filling is different from a half and the pre-quench state is not   insulating, the dynamical appearance of the topological edge states is visible already in a chain, it survives time averaging and can be observed also in the presence of chiral-breaking disorder and for instantaneous quenches. 
\end{abstract}


\maketitle

\section{Introduction}
The existence of topological edge states protected by some symmetry is perhaps  the most striking feature characterizing topological insulators and superconductors.  
Evidence of the  conducting channels in 2D topological insulators and of Dirac surface states in 3D Topological Insulators has been found in a number of experiments through transport, magnetotransport   and   photoemission spectroscopy measurements\cite{hasan-kane_review,zhang_review,ando_review,hasan-moore_review}. In  1D topological superconductors, the interpretation in terms of   Majorana quasi-particles\cite{alicea_review,aguado_review} of the    robust zero-bias peak observed in the conductance of spin-orbit coupled nanowires\cite{kouwenhoven_2012,kouwenhoven_2018} has led to some controversy\cite{frolov_2021}. A more direct evidence of these exotic quasi-particles seems to be provided by  spatially resolved  spectroscopic   techniques applied to ferromagnetic atom chains\cite{yazdani_2014}. In fact, in last years an increasing number of works have been devoted to the search for a real-space imaging of topological phases and states   \cite{yazdani_2014,molenkamp_PRX_2013,molenkamp_naturemat_2013,goldman_2013,morgenstern_2015,yu_2017,peregbarnea_2017,yoshino_2019,voigtlander_2021}.

After a decade characterized by a remarkable effort   to find  signatures of these edge states in various materials,  presently one of the most fascinating challenges  in Physics is the possibility to manipulate these states and to  possibly encode information therein\cite{zhang_PNAS,he-he_2019,guo_2019}.
To this purpose, the implementation with cold atoms in optical lattices\cite{dassarma_2008,bercioux,oh_2011,goldman_2012,bloch_2013,lee,lecheminant,ng,zoller,deng,orth,asboth-lewenstein,yang,zhu_2017,vishwanath,wu_2017,zhu_2018,an_2019} offers a twofold advantage, namely  a pretty reliable system isolation from the environmental decoherence, and an extremely precise control of the system Hamiltonian. In particular, it is possible to realize quantum quenches of the Hamiltonian parameters\cite{calabrese_2006,polkovnikov_review,eisert_2015,mitra_2018}, both over the entire system and on a spatially localized portion.  These experimental advances thus 
also bring up new interesting questions about topological systems. Consider, for instance, a topological insulator characterized by some symmetry and suppose that, by a quantum quench   preserving such symmetry, the system is dynamically brought from the trivial to the topological phase,  passing through a gap closing. Can one observe  the  topological states {\it dynamically} appear in real-space at  its  edges? Conversely, how do they evolve  and possibly disappear when the quench is towards the trivial phase?  In this paper we aim to answer  these questions, focussing on a prototypical  case, namely  the Su-Schrieffer-Heeger (SSH) model\cite{SSH_PRL1979,SSH_PRB1980}.\\

The SSH model  describes spinless  fermions in a bipartite one-dimensional lattice through the following tight-binding Hamiltonian 
\begin{eqnarray}\label{SSH-clean}
\hat{\mathcal{H}}_{SSH}&=&v \sum_{j}   \left( \hat{c}^{\dagger}_{j,A} \hat{c}^{}_{j,B}\,+\hat{c}^{\dagger}_{j,B} \hat{c}^{}_{j,A}\right) + \\
& & +w \sum_{j}    \left( \hat{c}^{\dagger}_{j,B} \hat{c}^{}_{j+1,A}\,+\hat{c}^{\dagger}_{j+1,A} \hat{c}^{}_{j,B}\right) \quad, \nonumber
\end{eqnarray}
where $\hat{c}^{\dagger}_{j,s}$ and $\hat{c}^{}_{j,s}$ denote the fermionic creation/annihilation operators for electrons localized at atom  $s=A,B$ within the $j$-th cell of the lattice,  whereas $v$ and $w$ indicate the intra- and inter-cell tunneling amplitudes, respectively.  The model, first introduced in the description of opto-electronic properties of polyacetylene\cite{SSH_PRL1979,SSH_PRB1980,barford_book},  is considered as a paradigmatic example of    one-dimensional topological insulators\cite{ungheresi_book,shen_book}. Indeed at half filling (one electron per unit cell) the SSH model  describes a band insulator characterized by a band gap $2\varepsilon_g$ with $\varepsilon_g=||v|-|w||$ and by a sublattice symmetry called chiral symmetry, which identifies   for $|v|<|w|$ and for $|v|>|w|$ two topologically different  phases that cannot be connected to each other without closing the gap.  In the topologically non-trivial phase, a  SSH chain exhibits  at its edges localized  states that are protected by the chiral symmetry.   {Recently,}    soliton states and topological indices of the SSH model have been experimentally observed in  implementations  with cold atoms\cite{gadway_2016,gadway_2019}.
Moreover, the effects of time-dependent perturbations to the SSH  hopping amplitudes have been analyzed in the context of  topologically protected quantum gates\cite{asboth_PRB_2016} and Floquet nonequilibrium states generated by periodic drives\cite{foatorres_PRA_2015,dutta_2019a,dutta_2019b,xie_2019}.   \\

Consider now a  SSH chain-lattice, initially in the ground state of the topologically trivial phase ($|v|>|w|$), at half filling, and perform
  a   quench of the hopping amplitudes to  the topological phase ($|v|<|w|$). If the quench remains within the chiral symmetry class, the gap closes at some time and, by  inspecting the dynamical evolution of the occupancy at each chain site, one would expect  localized topological states to gradually emerge at the chain edges. Here we show that this is  not the case: the site occupancy remains exactly equal to the one of the trivial pre-quench state at any time and any site, including at the chain edges, regardless of the quench protocol (fast or slow) and even in the presence of chiral disorder. As we shall demonstrate, the reason boils down to the charge conjugation  or time-reversal symmetries of the pre-quench state and of the quenching Hamiltonian. In an open chain, these additional symmetries are typically present and  completely mask any effect of the quench in real-space, including the appearance of the topological states. 
{Effects of a quench can be observed in real-space occupancies} only when such symmetries are broken,  which can be done in two ways: i)  remaining within the topological insulator framework, i.e.  preserving  the chiral symmetry and the half-filling condition; ii) by ``brute force", i.e. by breaking the chiral symmetry  and/or by moving away from an insulating state.  

We shall first explore the first option and propose two ways to observe the dynamical effects of the quench in real-space. The  quench protocols are based on a local quench, where  a ring lattice  is cut  into a  chain or, viceversa, two edges of a chain are bridged to form a ring. In both cases   the presence of a magnetic flux threading the ring is crucial to induce a real-space dynamical response to the quench, which is different depending on whether the involved chain is in the   trivial or in the topological phase.
  
Then, we shall explore the second option and analyze the effects of quenches beyond the framework of topological insulator, i.e. by breaking the chiral symmetry and by considering filling values different from $1/2$, where the SSH model describes a metallic state. We find that the optimal way to observe   the dynamical appearance of the edge states characterizing the topological insulator is to have a slightly metallic system. Then, the dynamical appearance of these states is robust even in the limit of short quench time and in the presence of chiral breaking disorder.

The paper is organized as follows. In Sec.\ref{sec-2}, after   briefly  summarizing   the aspects of the SSH model that are needed to illustrate our results, including its symmetries, we shall describe  the method we used to compute the dynamical evolution. In Sec.\ref{sec-3} we present a general theorem ensuring that the site occupancy remains locked to $1/2$ when charge conjugation symmetry is present. In particular, this explains  the case of a quenched half-filled SSH chain. Then, in Sec.\ref{sec-4} we  show how to violate the hypotheses of the theorem  and observe real-space effects of the quench without breaking the chiral symmetry.
Finally, after analyzing the effects of chiral breaking terms and of a filling different from $1/2$ in Sec.\ref{sec-5}, we discuss  
 our results and draw our conclusions  in Sec.\ref{sec-6}.

\section{Model, symmetries and method}
\label{sec-2}
%%%%%%%%%%%%%%%%%%%%%%%
%%%%.   Generalized SSH model.    %%%% 
%%%%%%%%%%%%%%%%%%%%%%%
\subsection{Generalized SSH model and symmetries}
\label{sec-symm}
In this paper we  consider a generalized SSH model
\begin{eqnarray}\label{SSH-gen}
\hat{\mathcal{H}}_{SSH,\chi} &=&\sum_{j} \left( v^{}_j  \hat{c}^{\dagger}_{j,A} \hat{c}^{}_{j,B}\,+v_j^* \hat{c}^{\dagger}_{j,B} \hat{c}^{}_{j,A} \right)\, +\\
& & + \, \sum_{j} \left( w^{}_j \hat{c}^{\dagger}_{j,B} \hat{c}^{}_{j+1,A}\,+w_j^* \, \hat{c}^{\dagger}_{j+1,A} \hat{c}^{}_{j,B} \right)  \nonumber
\end{eqnarray}
 extending the SSH Hamiltonian (\ref{SSH-clean}) to the case where the tunneling amplitudes $\{ v_j\,,w_j\}$ are  possibly complex and site-dependent.  Here $j=1,2\ldots, M$, where $M$ is the number of cells in the lattice.
%%%%%%%%%%%%%%%%%%%%
%%%%     Symmetries    %%%
%%%%%%%%%%%%%%%%%%%%
Furthermore, because symmetries play an important role in the dynamical effects that we aim to discuss, it is worth briefly recalling the behavior of the Hamiltonian (\ref{SSH-gen})  under  three  transformations that are local on the lattice site operators. 
The first one is   {\it charge-conjugation}    $\mathcal{C}$, a linear and unitary transformation  mapping   the lattice site creation/annihilation operators as follows
\begin{equation}\label{C-def}
\left\{ \begin{array}{lcl}
\mathcal{C} \hat{c}^{}_{j,A}\mathcal{C}^{-1}  &=& \hat{c}^{\dagger}_{j,A} \\
\mathcal{C} \hat{c}^{}_{j,B}\mathcal{C}^{-1}  &=& -\hat{c}^{\dagger}_{j,B} 
\end{array}\right. \quad,
\end{equation}
and fulfilling $\mathcal{C}^{-1}=\mathcal{C}^{\dagger}=\mathcal{C}$.  The second one is the  {\it chiral transformation}  $\mathcal{S}$. Despite acting on the   lattice site operators in  the same way as  $\mathcal{C}$  
\begin{equation}\label{S-def}
\left\{ \begin{array}{lcl}
\mathcal{S} \hat{c}^{}_{j,A}\mathcal{S}^{-1}  &=& \hat{c}^{\dagger}_{j,A} \\
\mathcal{S} \hat{c}^{}_{j,B}\mathcal{S}^{-1}  &=& -\hat{c}^{\dagger}_{j,B} 
\end{array}\right.\quad,
\end{equation}
it is by definition anti-linear ($\mathcal{S}\,i\, \mathcal{S}=-i$) and anti-unitary $\langle \mathcal{S}\Psi_1 | \mathcal{S}\Psi_2\rangle=\langle  \Psi_1 | \Psi_2\rangle^*$. 
Finally, the {\it time-reversal}  transformation, which leaves  lattice site operators unaltered
\begin{equation}\label{T-def}
\left\{ \begin{array}{lcl}
\mathcal{T} \hat{c}^{}_{j,A}\mathcal{T}^{-1}  &=& \hat{c}^{}_{j,A} \\
\mathcal{T} \hat{c}^{}_{j,B}\mathcal{T}^{-1}  &=& \hat{c}^{}_{j,B} 
\end{array}\right.
\end{equation}
but is also anti-linear and anti-unitary, with $\mathcal{T}^2=\mathbb{I}$, as is the case for spinless fermions.
In fact, only two of these transformations are independent because the chiral symmetry $\mathcal{S}$ can be obtained as the product $\mathcal{S}=\mathcal{T} \, \mathcal{C}$.

The Hamiltonian (\ref{SSH-gen})  exhibits  the chiral symmetry~(\ref{S-def})
\begin{equation}\label{chiral-symm}
\mathcal{S} \hat{\mathcal{H}}_{SSH,\chi}\,\mathcal{S}^{-1} =\hat{\mathcal{H}}_{SSH,\chi} \hspace{0.5cm} \Leftrightarrow \hspace{0.5cm} \left[\hat{\mathcal{H}}_{SSH,\chi}\,,\,\mathcal{S} \right]=0\quad,
\end{equation}
and the  subscript $\chi$  stands in fact for `chiral'. In chiral-symmetric models like $\hat{\mathcal{H}}_{SSH,\chi}$, $\mathcal{T}$ and  $\mathcal{C}$ are intimately related. Indeed, because $\mathcal{C}=\mathcal{T} \, \mathcal{S}$, time-reversal and charge conjugation  transformations are either both preserved or both broken. In particular, for $\{ v_j \,,\, w_j\} \in \mathbb{R}$, the Hamiltonian (\ref{SSH-gen}) also commutes with time-reversal~$\mathcal{T}$ and charge conjugation $\mathcal{C}$. However, when $v_j=|v_j| e^{i \phi^v_j}$ and $w_j=|w_j| e^{i \phi^w_j}$ have non-vanishing complex phases, the preservation of $\mathcal{T}$ and $\mathcal{C}$ heavily depends on the geometric boundary conditions. 
  In particular, in a chain, i.e. a lattice with open boundary conditions (OBCs), $\mathcal{T}$ and $\mathcal{C}$ are always preserved, since such complex phases        can be eliminated through a canonical transformation onto the lattice operators $c_{j,A}\rightarrow  \tilde{c}_{j,A}=e^{i\alpha_j}c_{j,A}$ and $c_{j,B}\rightarrow  \tilde{c}_{j,B}=e^{i(\alpha_j+\phi^v_j)}c_{j,B}$, where $\alpha_1=0$ and $\alpha_j=\sum_{i=1}^{j-1} (\phi_i^v +\phi_i^w)$ for $j=2,\ldots M$, recasting the Hamiltonian into the case of real and positive tunneling amplitudes\cite{ungheresi_book}.   In contrast,   in a ring-shaped lattice, the periodic boundary conditions (PBCs) prevent the elimination of the phases of $v_j$ and $w_j$.
Physically, this can be understood in terms of the Peierls substitution\cite{peierls,graf-vogl}, where the complex phases of the tunneling amplitudes describe  the integral of a vector potential from one lattice site to next one. While in a chain  the vector potential can always be gauged out, in a ring this is not possible, for its circulation yields the magnetic flux $\Phi$ threading the ring, and one has $\sum_{j=1}^M (\phi^v_j+\phi^w_j) = 2\pi \Phi /\Phi_0 \neq 0$, where $\Phi_0=h/ {\rm e}$ is the flux quantum. Thus, in the SSH ring with a flux $\Phi \neq p \Phi_0/2$ (with $p \in \mathbb{Z}$), $\mathcal{T}$ and  $\mathcal{C}$  are broken.  \\

To perform our time-dependent analysis,  we represent  the second-quantized   Hamiltonian~(\ref{SSH-gen}) in the real-space basis as follows
\begin{equation} \label{H-2nd-1st}
\hat{\mathcal{H}}_{SSH,\chi} =\sum_{j_1,j_2=1}^{M} \sum_{s_1,s_2=A,B}^{}  \hat{c}^{\dagger}_{j_1, s_1} {H}^{}_{j_1 s_1, j_2 s_2}  \,\hat{c}^{}_{j_2, s_2}\quad,
\end{equation}
where
\begin{equation}
{H}= \left( \begin{array}{|cc|cc|cc|cc|cc|} 
0 & v_1 &  &  & & & & & & w^*_M \\
v_1^* & 0 & w_1 &   & & & & & &\\ \hline
  &   w^*_1 & 0 & v_2 & & & & & &\\ 
  &    & v_2^* & 0& w_2 & & & & &\\ \hline
  & & & w_2^*  & \ddots &   & &  & &\\ 
  &   & & &  &  \ddots  & & &  &\\ \hline
  & & & & &   & \ddots &     & &\\ 
  &   & & & & &  &  \ddots   & w_{M-1} &\\ \hline
  &     &   &   & & & &w^*_{M-1}& 0& v_M \\ 
w_M  &    &   &  & & & & & v_M^* & 0\\ \hline
\end{array} \right)\label{SSH-mat}
\end{equation}
is the related first-quantized Hamiltonian matrix, whose entries ${H}^{}_{j_1 s_1, j_2 s_2}$
are labelled by  the cell $j$  and the site $s=A,B=+/-$  within the cell.   In terms of the first quantized Hamiltonian  (\ref{SSH-mat}), symmetries are expressed in a different way as compared to the second quantized Hamiltonian (\ref{SSH-gen}). Explicitly, the
  chiral symmetry (\ref{chiral-symm}) implies
\begin{equation} \label{S-def-mat}
S \, H\, S^{-1} =-H \hspace{0.5cm} \Leftrightarrow \hspace{0.5cm} \{ H, S \} =0
\end{equation}
where $S=\oplus_{j=1}^M (\sigma_z)_j$ is the first-quantized version of the chiral transformation $\mathcal{S}$ defined in {Eq.}(\ref{S-def}), and is unitary.  
From {Eq.}(\ref{S-def-mat}) one straightforwardly deduces that, for any realization of the parameters $\{ v_j \,,\, w_j\}$,  the single-particle spectrum  is symmetric around $\varepsilon=0$. Indeed if $\psi$ is a single-particle wavefunction with eigenvalue $\varepsilon$, i.e. ${H}\psi =\varepsilon \, \psi$, the wavefunction $S\psi$, obtained from $\psi$ by changing the sign at the $B$-sites, is also an eigenfunction of ${H}$ with eigenvalue $-\varepsilon$.  The  set of eigenfunctions of {Eq.}(\ref{SSH-mat}) can thus be chosen as $\{ \psi_{\alpha} \}$ (positive eigenvalues $\varepsilon_\alpha>0$) and $\{ S\psi_\alpha \}$ (negative eigenvalues $-\varepsilon_\alpha<0$), where $\alpha=1,\ldots M$ is the quantum number running over the positive spectrum. 
The corresponding operators 
\begin{equation}\label{gamma-def}
\left\{ \begin{array}{lcl}
\gamma_{\alpha,+}= \displaystyle \sum_{j=1}^{{M}} \sum_{s=A/B=\pm} (\psi^*_\alpha)_{j,s}\, \hat{c}_{j,s} \\
\gamma_{\alpha,-}= \displaystyle  \sum_{j=1}^{{M}} \sum_{s=A/B=\pm} (\psi^*_\alpha)_{j,s}\, (-1)^s \, \hat{c}_{j,s}
\end{array}
\right.
\end{equation}
diagonalize the Hamiltonian (\ref{SSH-gen})
\begin{equation}
\hat{\mathcal{H}}_{SSH,\chi} = \sum_{\alpha} \varepsilon_\alpha \left( \gamma^\dagger_{\alpha,+}\gamma^{}_{\alpha,+}-\gamma^\dagger_{\alpha,-}\gamma^{}_{\alpha,-}\right)
\end{equation}
and fulfill the relations
\begin{equation}\label{gamma-prop}
\mathcal{S} \gamma^{}_{\alpha,\pm} \mathcal{S}^{-1}= \gamma^{\dagger}_{\alpha,\mp}
\end{equation}

While the chiral symmetry (\ref{S-def-mat}) always holds for {Eq.}(\ref{SSH-mat}), time-reversal and charge-conjugation symmetries hold  if the Hamiltonian ${H}$ fulfills further properties. Specifically the former symmetry holds if ${H}$ is real
\begin{equation} \label{TRS-mat}
T \, H\, T^{-1} =H^{*}=H
\end{equation}
where $T=K$ denotes the complex conjugation and is anti-unitary, whereas the latter symmetry holds if
\begin{equation} \label{CCS-mat}
C \, H\, C^{-1} =S\,{H}^*\, S^{-1}= -H
\end{equation}
where ${C}=S \,T$  is the first-quantized version of  $\mathcal{C}$ [see {Eq.}(\ref{C-def})] and is   anti-unitary.

Note that the  hopping amplitude $w_M$ appearing in the lower-left and upper-right corners of {Eq.}(\ref{SSH-mat}) is vanishing for a chain. In such a case,  an argument similar to the one used above for the second quantized Hamiltonian, leads to conclude  that Eqs.(\ref{TRS-mat}) and (\ref{CCS-mat}) always hold, as can be checked by   merely redefining  the real-space basis by   local phase factors.

We conclude this subsection by recalling that, for {\it homogeneous} hopping amplitudes ($v_j\equiv v$ and $w_j\equiv w$) the model (\ref{SSH-gen}) can be exactly solved both in a ring (PBCs) and in a chain (OBCs). In particular, in the ring geometry and in the thermodynamic limit one can identify two different topological classes\cite{ungheresi_book}, and by analyzing the chain one can {see} that one phase is topologically non-trivial, hosting two discrete levels in the spectrum near  $\varepsilon=0$, which correspond to states localized at the edges. For the sake of completeness, a short summary  of these aspects in given in the Appendix.
 
%%%%%%%%%%%%%%%%%%%%%%%
%%%%.  Method    %%%% 
%%%%%%%%%%%%%%%%%%%%%%%
\subsection{Quenches, density matrix approach and observables}
\label{sec-method}
%In the following, we shall investigate the dynamical effects of a quench in the parameters $\{ v_j , w_j\}$  of the Hamiltonian~(\ref{SSH-gen}). Specifically, during a  time lapse $\tau_q$ starting at a time $t_0$ and terminating at time $t_f$, the Hamiltonian  $\hat{\mathcal{H}}_{SSH,\chi}(t)$ varies in time from a pre-quench  $\hat{\mathcal{H}}^{\rm pre}=\hat{\mathcal{H}}_{SSH,\chi}(t<t_0)$ to a post-quench Hamiltonian $\hat{\mathcal{H}}^{\rm post}=\hat{\mathcal{H}}_{SSH,\chi}(t>t_f)$. The quench protocol specifies the way the parameters $\{ v_{j}(t), w_j(t)\}$ in {Eq.}(\ref{SSH-gen}) are varied during the quench time $\tau_q=t_f-t_0$. 

In the following, we shall investigate the dynamical effects of a quench in the parameters $\{ v_j , w_j\}$  of the Hamiltonian~(\ref{SSH-gen}). Specifically, the system is prepared in an initial state~$\boldsymbol\rho^{\rm pre}$, typically the ground state or the thermal equilibrium state of a pre-quench Hamiltonian $\hat{\mathcal{H}}^{\rm pre}=\hat{\mathcal{H}}_{SSH,\chi}(t<t_0)$. Then, at $t=t_0$ the system is disconnected from the environment and  the dynamics is unitarily governed by the Hamiltonian  $\hat{\mathcal{H}}_{SSH,\chi}(t)$, which  varies  until a time $t_f$ from   $\hat{\mathcal{H}}^{\rm pre}$ to a post-quench Hamiltonian $\hat{\mathcal{H}}^{\rm post}=\hat{\mathcal{H}}_{SSH,\chi}(t>t_f)$. The quench protocol specifies the way the parameters $\{ v_{j}(t), w_j(t)\}$ in {Eq.}(\ref{SSH-gen}) are varied during the quench time $\tau_q=t_f-t_0$.

Although in our analysis we shall mainly focus on short  quench time limit ($\tau_q\rightarrow 0$), we shall keep the parameter time dependence arbitrary because, as we shall see, some results are independent of the specific quench protocol.
Moreover, we shall deal with both global and local quenches. A global quench involves a change in a significant number (scaling like the number $M$ of cells) of hopping amplitudes along the chain. This occurs, for instance, when all the   hopping amplitudes of a homogeneous chain ($v_j\equiv v$ and $w_j\equiv w$) are brought from the  trivial to the  topological phase. In contrast, a local quench only  involves a limited number of hopping amplitudes. For instance, the cutting  of a ring into a chain is described by quenching to zero the  hopping amplitude of one single bond.  Note that, because of {Eq.}(\ref{chiral-symm}), the chiral symmetry is preserved at any time, so that the quench occurs within the chiral symmetry class. Yet, the result of a quench depends not only on the quenching Hamiltonian, but also on the pre-quench state and its symmetries, as we shall see. Furthermore, in Sec.\ref{sec-5}, we shall also analyze  the quench in chiral symmetry broken cases.\\

We shall be interested in one-body observables $\hat{\mathcal{A}}=\sum_{j_1 s_1, j_2 s_2} \hat{c}^{\dagger}_{j_1, s_1} {A}^{}_{j_1 s_1, j_2 s_2} \,\hat{c}^{}_{j_2, s_2}$, whose expectation values are straightforwardly evaluated in terms of the single-particle density matrix $\rho_{j_1 s_1, j_2 s_2}(t)={\rm Tr} \{ c^\dagger_{j_2, s_2} \hat{c}^{}_{j_1 s_1} \hat{\boldsymbol{\rho}}(t)\}$, where $\hat{\boldsymbol{\rho}}(t)$ denotes the dynamical evolution of the full system density matrix and ${\rm Tr}$  the trace over the  Fock space. Due to the quadratic structure of {Eq.}(\ref{H-2nd-1st}), the Liouville-von Neumann equation for $\hat{\boldsymbol{\rho}}$  straightforwardly implies the dynamical equation for $\rho$, which reads
\begin{equation}\label{LvN-mat}
i\hbar \frac{d \rho}{dt} =\left[ H(t), \rho \right] \quad.
\end{equation}
We numerically solve  {Eq.}(\ref{LvN-mat}) with the initial condition $\rho(t_0)=\rho^{\rm pre}$ corresponding to the single-particle density matrix of the  pre-quench state, typically the ground state of the pre-quench Hamiltonian. Then, the expectation values of an observable $\hat{\mathcal{A}}$ are  obtained as
\begin{equation}
\langle \hat{\mathcal{A}}\rangle(t)= {\rm tr} \left\{ A \, \rho(t) \right\}
\end{equation}
where ``${\rm tr}$" denotes the trace over the single-particle Hilbert space.
In particular,  {we shall henceforth focus on the site occupancy,} evaluated as 
\begin{equation}
N_{j,s}(t)=\langle \hat{n}_{j,s}\rangle(t) = \rho_{j s,js}(t) \quad,
\end{equation}
{and on  the cell polarization } 
\begin{equation}
P_j(t)=\langle \hat{n}_{j,A}\rangle(t)-\langle\hat{n}_{j,B}\rangle(t) \quad,
\end{equation}
{obtained} as $P_j=\rho_{j A,jA}-\rho_{j B,jB}$. The total number of electrons $N_e=\sum_{j,s}N_{j,s}$ is simply given by $N_e={\rm tr}\rho$ and is constant as a consequence of {Eq.}(\ref{LvN-mat}).
In Sec.\ref{sec-5} we shall also discuss the non-equilibrium energy distribution in the post-quench eigenbasis $\{ \lambda \}$, where $\hat{\mathcal{H}}^{\rm post}=\sum_\lambda  \varepsilon_{\lambda}  \hat{n}_{\lambda} $ is diagonal. The  energy distribution is obtained as 
\begin{equation}
\langle \hat{n}_{\lambda} \rangle(t)=\sum_{j_1,j_2=1}^M \sum_{s_1, s_2=A,B} U_{\lambda , j_2 s_2} U^*_{\lambda  , j_1 s_1} \, \rho_{j_2 s_2, j_1 s_1}(t)
\end{equation}
where  $U_{\lambda , j  s }=\langle\lambda | j  s  \rangle$ is the unitary matrix determining the single-particle change of basis from the real-space basis to the post quench eigenbasis.
%\begin{equation}
%\begin{pmatrix}
%  \begin{matrix}
%  a & b \\
%  c & d
%  \end{matrix}
%  & \rvline & \bigzero \\
%\hline
%  \bigzero & \rvline &
%  \begin{matrix}
%  a & b \\
%  c & d
%  \end{matrix}
%\end{pmatrix}
%\end{equation}

%%%%%%%%%%%%%%%%%%%%%%%%%%%%%%%%%%%%%%%
%%%%%%%%%%%%%%%%%%%%%%%%%%%%%%%%%%%%%%%
%%%%%%%%%%%%%%%%%%%%%%%%%%%%%%%%%%%%%%%
\section{Quenches in   half-filled SSH models: The locking of site occupancy}
\label{sec-3}
\begin{figure}[h]
\centering
\includegraphics[width=0.8\linewidth]{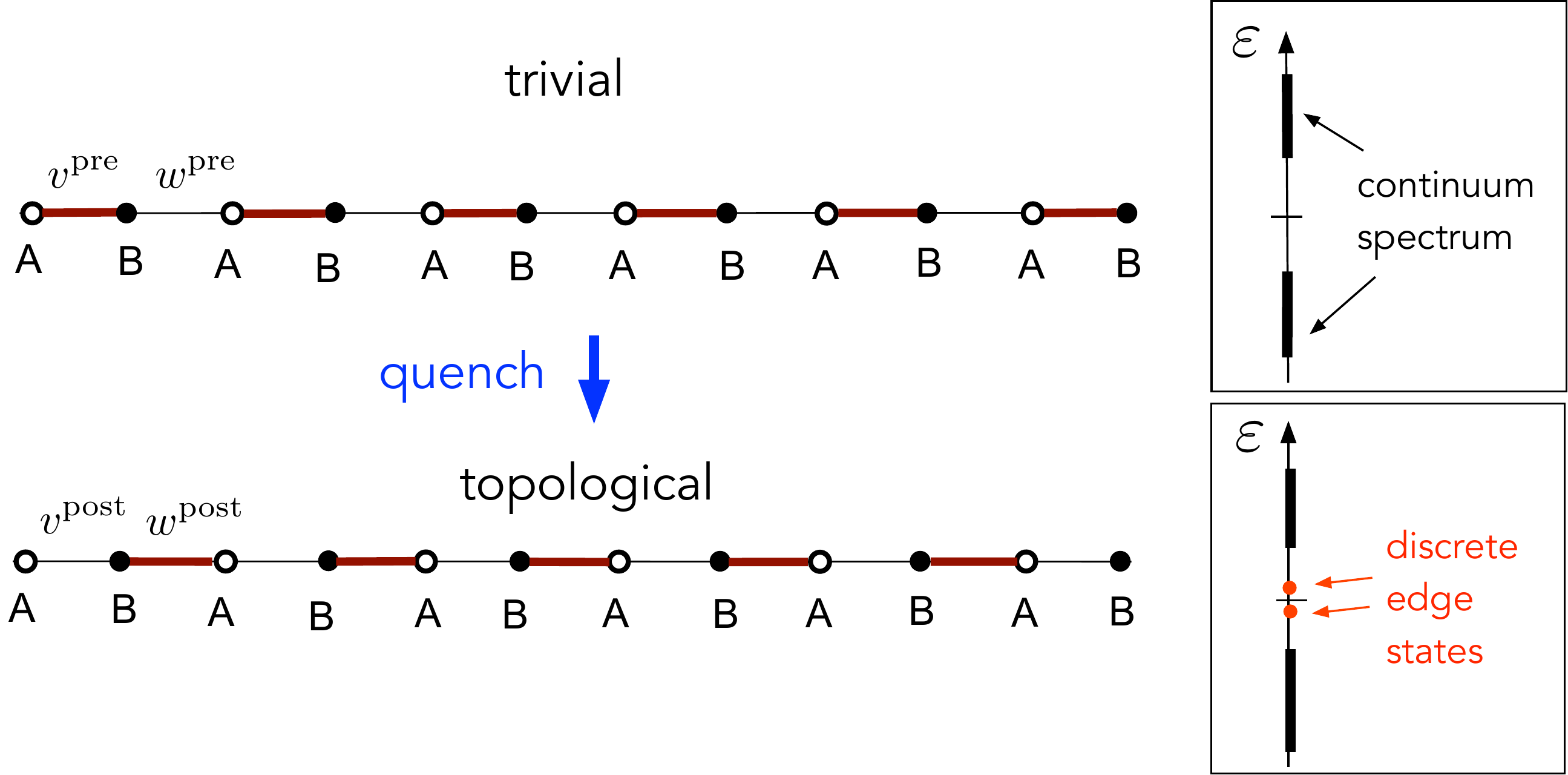}
\caption{\label{Fig1-chain-quench-scheme} A global quench is applied to a half-filled SSH chain from the trivial to the topological phase. The insets on the right-hand side sketch the spectra of the two phases: while the trivial phase only exhibits a continuum spectrum, the topological phase also hosts two discrete states. }
\end{figure}

We start by considering a chain of the customary SSH model  {Eq.}(\ref{SSH-clean}) with homogeneous hopping amplitudes, which can be assumed to be positive ($v,w>0$).  Let the pre-quench state be the half-filled  ground state   of the chain  in the trivial phase, so that   there is one electron per cell ($N_e=M$), i.e. half an electron per site on average, and
\begin{equation}
 w^{\rm pre} < v^{\rm pre}  \quad,
\end{equation}
where $w^{\rm pre}\equiv w(t< 0)$ and $v^{\rm pre}\equiv v(t< 0)$.
No edge state is present.  At $t=0$ we start to quench  the Hamiltonian parameters towards  the topological phase, as sketched in Fig.\ref{Fig1-chain-quench-scheme}. This means that, within a quench time $\tau_q$, the values  of the hopping amplitudes are brought to
\begin{equation}
 w^{\rm post} > v^{\rm post} \quad,
\end{equation}
where $w^{\rm post} \equiv w(t >\tau_q)$ and  $v^{\rm post} \equiv v(t >\tau_q)$.
%%%%%%%%%%%%%%%%%%%%%%
%%%%%%%%%%%%%%%%%%%%%

At first, one would expect the discrete states characterizing the   chain spectrum in the topological phase to gradually appear in real-space, causing an occupancy increase localized at the two edges. However, this is not the case: we find that the site occupancy  is  locked at $1/2$ {\it at any time} and {\it at any site}, including the chain edges 
\begin{equation}\label{locking}
N_{j,s}(t)=1/2 \hspace{2cm} \forall t \hspace{0.5cm} \forall j,s
\end{equation}
just like in the trivial  pre-quench phase. Notably, such a locking of the real-space occupancy   occurs   
 for any quench duration $\tau_q$,  regardless of the specific way one changes the hopping amplitudes from $(v^{\rm pre},w^{\rm pre})$ to   $(v^{\rm post},w^{\rm post})$. Furthermore, it also holds in the presence of chiral disorder and/or if the pre-quench state is a thermal state at finite temperature.   Indeed  the result  (\ref{locking}) is a consequence of a   general theorem that we shall  prove here below. {Before doing that,} it is worth emphasizing that  the quench does affect the system, though. For instance, the energy distribution of the post-quench Hamiltonian  strongly differs from the pre-quench equilibrium distribution and exhibits a striking band population inversion, as has been proven in Ref.\cite{sassetti_PRB2018} for a  SSH ring exposed to a sudden quench.

\subsection{General theorem about site occupancy}
\label{sec-theorem}
The   following general result  can be proven: (i) If the pre-quench state ($t =t_0$) is invariant under charge-conjugation
\begin{equation}\label{cond-rho}
\mathcal{C} \hat{\boldsymbol{\rho}}^{\rm pre}\mathcal{C}^{-1} = \hat{\boldsymbol{\rho}}^{\rm pre}
\end{equation}
and (ii) if the time-dependent Hamiltonian $\hat{\mathcal{H}}(t)$ characterizing the quench ($t>t_0$) commutes   with charge-conjugation  transformation   
\begin{equation}\label{cond-Ham}
\left[ \hat{\mathcal{H}}_{SSH,\chi}(t > t_0)\,,\, \mathcal{C} \right] =0  \quad,
\end{equation}
then {Eq.}(\ref{locking}) holds. The proof starts by recalling that the pre-quench state $\hat{\boldsymbol{\rho}}^{\rm pre}$ evolves as $\hat{\boldsymbol{\rho}}(t)= \mathcal{U}(t) \hat{\boldsymbol{\rho}}^{\rm pre}  \mathcal{U}^\dagger(t)$, where  the  evolution operator is
\begin{equation}
\mathcal{U}(t)=\overleftarrow{\rm T}\left[ \exp\left(-\frac{i}{\hbar} \int_{t_0}^t \hat{\mathcal{H}}(t^\prime) dt^\prime \right)\right] \hspace{2cm} \forall t > t_0
\end{equation}
and   $\overleftarrow{\rm T}$ denotes the time-ordering. Moreover, the property (\ref{cond-Ham})  and the linearity of $\mathcal{C}$ imply that
\begin{equation}\label{comm-U}
\left[ \mathcal{C}\, ,\, \mathcal{U}(t)\right]=0 \hspace{2cm} \forall t>t_0
\end{equation}
By using Eqs.(\ref{cond-rho}) and (\ref{comm-U}) the time evolution of the    site occupancy $N_{j,s}=\langle \hat{n}_{j,s}\rangle$ is then computed as
\begin{eqnarray}
N_{j,s}(t)&=&{\rm Tr}\left\{ \hat{\boldsymbol{\rho}}(t) \, \hat{n}_{j,s}\right\} = \nonumber \\
&=&{\rm Tr}\left\{  \mathcal{U}(t) \hat{\boldsymbol{\rho}}^{\rm pre}  \mathcal{U}^\dagger(t)\, \hat{n}_{j,s}\right\} = \nonumber \\
&=&{\rm Tr}\left\{  \mathcal{U}(t) \,\mathcal{C} \hat{\boldsymbol{\rho}}^{\rm pre}\mathcal{C}^{-1}   \mathcal{U}^\dagger(t)\, \hat{n}_{j,s}\right\} = \nonumber \\
&=&{\rm Tr}\left\{ \mathcal{C} \mathcal{U}(t) \hat{\boldsymbol{\rho}}^{\rm pre} \, \mathcal{U}^\dagger(t)\, \mathcal{C}^{-1}  \hat{n}_{j,s}\right\} = \nonumber \\
&=&{\rm Tr}\left\{ \mathcal{U}(t)  \hat{\boldsymbol{\rho}}^{\rm pre} \, \mathcal{U}^\dagger(t)\, \mathcal{C}  \hat{n}_{j,s}  \mathcal{C}^{-1}\right\} = \nonumber \\
&=&{\rm Tr}\left\{ \hat{\boldsymbol{\rho}}(t)\,  (1-\hat{n}_{j,s} )\right\} = \nonumber \\
&=& 1-N_{j,s}(t) \label{proof}
\end{eqnarray}
where we have used $\mathcal{C}=\mathcal{C}^{-1}$ and $\mathcal{C}  \hat{n}_{j,s}  \mathcal{C}^{-1}=1- \hat{n}_{j,s} $. The result {Eq.}(\ref{locking}) follows from {Eq.}(\ref{proof}), and shows that  the site occupancy remains locked to its trivial phase value $1/2$.
We also observe that, by a very similar argument, the   hypotheses of the theorem also imply   that the off-diagonal single-particle density matrix entries are always either real or purely imaginary, at any time. Specifically $\rho_{i A, jB}(t)=\langle c^{\dagger}_{j B} c^{}_{i A} \rangle$ is real $\forall i,j$, while $\rho_{i s, js}(t)=\langle c^{\dagger}_{j s} c^{}_{i s} \rangle$ is purely imaginary   $\forall i\neq j$ and $s=A,B$.

\subsection{Global quench in a SSH chain}
We shall now show that a quench of the half-filled SSH chain satisfies the hypotheses Eqs.(\ref{cond-rho}) and (\ref{cond-Ham}) of the above theorem, whence one  straightforwardly deduces    the locking of the site occupancy, {Eq.}(\ref{locking}).  Indeed {Eq.}(\ref{cond-Ham}) is satisfied  by $\hat{\mathcal{H}}_{SSH,\chi}(t)$ because,  as observed in Sec.\ref{sec-symm}, in a chain with OBCs the  SSH model {Eq.}(\ref{SSH-gen}) preserves both charge-conjugation $\mathcal{C}$ and 
 time-reversal symmetry  $\mathcal{T}$. Furthermore, if the pre-quench state is the thermal equilibrium state at half-filling ($\mu=0$) of the pre-quench  SSH Hamiltonian  $\hat{\mathcal{H}}^{\rm pre}$
\begin{equation}\label{rho-pre-thermal}
\hat{\boldsymbol{\rho}}^{\rm pre} = \frac{e^{-\beta \hat{\mathcal{H}}^{\rm pre}}}{{\rm Tr}[e^{-\beta\hat{\mathcal{H}}^{\rm pre}}]}
\end{equation}
where $\beta=1/k_B T$ is the inverse temperature, the symmetry $[ \hat{\mathcal{H}}^{\rm pre} ,\, \mathcal{C} ] =0$ straightforwardly implies {Eq.}(\ref{cond-rho}). In particular, this is true for the half-filled ground state
$
\hat{\boldsymbol{\rho}}^{\rm pre} =| {\rm H.F.} \rangle \langle {\rm H.F.} |
$, where $| {\rm H.F.} \rangle$   is constructed by occupying all the negative energy states of $\hat{\mathcal{H}}^{\rm pre}$
$
| {\rm H.F.} \rangle = \prod_{\alpha=1}^M \gamma^{\dagger}_{\alpha,-} |0\rangle
$ 
and is non-degenerate.\\

This explains why in a half-filled SSH chain the quench does not lead to any change of the site occupancy, which remains uniform and constant regardless of i) the specific quench time and protocol, ii) the presence of chiral disorder in the tunneling amplitudes $\{ v_j,w_j\}$, and iii) finite temperature of the pre-quench thermal state. In particular, in a chain it is impossible to observe the appearance of the topological states or any other difference  between the trivial and the topological phase in real-space occupancies. In Sec.\ref{sec-4} we shall propose a different setup where real-space effects of a quench can be observed. However, we wish to first provide a more physical justification for the result {Eq.}(\ref{locking}).

\subsubsection{The case of infinitely slow quench:  Comparison between the trivial and topological half-filled ground states.}
\label{sec-comparison}
\begin{figure}[h]
\centering
\includegraphics[width=0.75\linewidth]{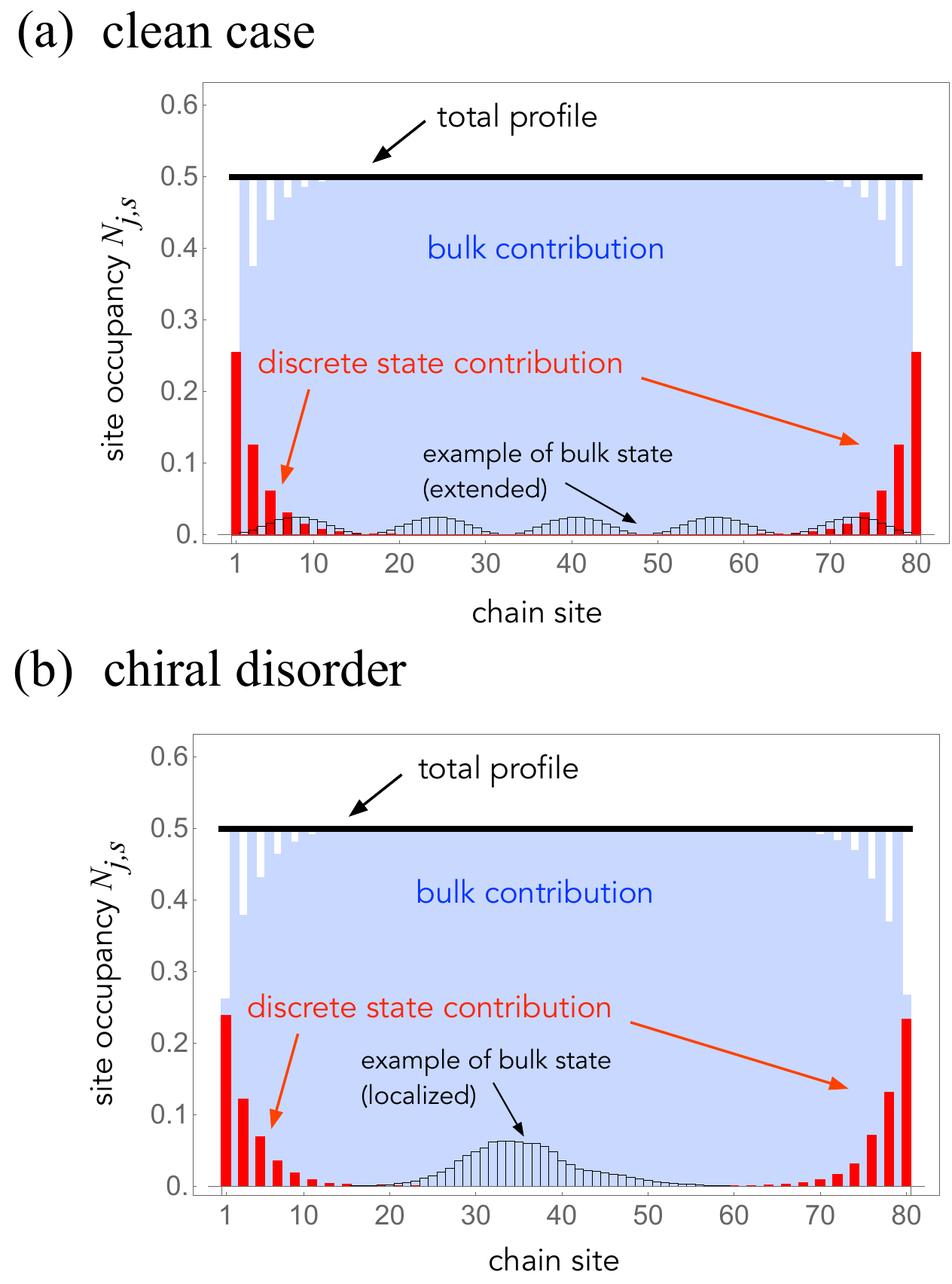}
\caption{\label{Fig2-chain-groundstate-density-profile} The  site occupancy profile of the half-filled SSH chain (${N}=80$ sites, i.e. $M=40$ cells) in the topological phase  (thick black curve), the discrete level contribution  (edge states, in red) and the contribution of all the occupied states in the continuum spectrum (bulk states, blue). (a) in the clean case, where the tunneling amplitudes are homogeneous  ($v_j \equiv v$, $w_j \equiv w$, with $v=0.7 w$), all bulk states are extended, and the thin black curve shows an example of a bulk state; (b) in the chiral-disordered case, where the tunneling amplitudes $v_j, w_j$ are random variables with average values fulfilling $v=0.7w$ and with disorder strength $d=0.1$, the bulk states    are also localized, and one example is shown by the thin black curve. In both cases the edge and bulk states contributions are perfectly complementary, so that the total site occupancy profile is flat and equal to $1/2$ everywhere.}
\end{figure}

Because the result {Eq.}(\ref{locking})  is valid  for any quench protocol, it holds in particular for  an infinitely slow quench ($\tau_q\rightarrow \infty$), where the pre-quench ground state evolves into the post-quench ground state. In this particular limit,   the result {Eq.}(\ref{locking}) can thus be understood by comparing the site occupancy profile of the trivial and topological half-filled ground states.
   In the trivial phase, where the spectrum is purely continuum, the uniform pattern $N_{j,s}^{\rm pre}\equiv 1/2$ is expected from the contribution of the  bulk states   extending over the entire chain.
In the  topological phase, where the additional discrete levels  $\pm \varepsilon^{\rm edge}$ near $\varepsilon=0$ are present, the site occupancy profile results from two types of contributions.
The red curve in Fig.\ref{Fig2-chain-groundstate-density-profile}(a) shows the discrete state contribution  localized at the chain edges [see  {Eq.}(\ref{eigenvec-edge-chain})], while the thin black curve displays, as an illustrative example, the contribution of one bulk state, whose wavefunction extends over the entire chain [see {Eq.}(\ref{eigenvec-bulk-chain})]. 
Notably, the  blue curve, describing the contribution of {\it all} the occupied bulk states  of the chain,  features two dips at the edges, which are perfectly complementary to  the edge state contribution: The bulk states  ``feel" the presence of the edge states and  make room for them by slightly modifying  their behavior  near the boundaries  with respect to the trivial phase. This can be considered as a real-space imaging of the bulk-boundary correspondence. The thick black curve  is the sum of the two contributions and uniformly takes the value $N_{j,s}^{\rm post} \equiv 1/2$. Thus the half-filled ground state of the chain in the topological phase  does not show any different feature in real-space occupancy with respect to the trivial phase, despite the presence of the edge states in the spectrum. 
%%%%%%%%%%%%%%%%%%%%%%
%%%%%%%%%%%%%%%%%%%%%

Such a lack of difference seems at first   to contradict the argument that is customarily invoked to illustrate the emergence of the edge states   in the topological phase, based on the dimerized limit of the chain: When the extremal links of the chain are very weak, $v/w\rightarrow 0$, the outmost chain sites host a localized electron. However, this can only hold when the number $N_e$ of electrons in the chain is $M+1$. At half filling, $N_e=M$, only one electron can be accommodated in the two edge sites and, in fact, each of them hosts ``half an electron".  Indeed the  red curve of Fig.\ref{Fig2-chain-groundstate-density-profile},  peaked at {\it both} chain edges,  describes   the contribution of only {\it one} discrete state, namely the one  at energy $-\varepsilon^{\rm edge}$, which is occupied   in the half-filled ground state.

We also emphasize that such uniform site occupancy profile is not merely due to the  accidental spatial parity\cite{nota-parity} of the homogeneous SSH model (\ref{SSH-clean}). The very chiral symmetry   forbids disorder to localize the two discrete states on opposite sides of the chain: The two wavefunctions $\psi^{\rm edge}_{\pm}$  with opposite energies $\pm \varepsilon^{\rm edge}$ are mapped   into each other  by a mere sign change  in the $B$-sites through the chiral transformation $S$~[see Sec.\ref{sec-symm}], so that their square moduli have to coincide, even in the presence of chiral disorder.
This is illustrated in Fig.\ref{Fig2-chain-groundstate-density-profile}(b), which refers to a disordered SSH model realized by taking $w_j=w(1+\xi_j\, d)$ and $v_j=v(1+\eta_j \, d)$ in {Eq.}(\ref{SSH-gen}), where $\{\xi_j,\eta_j\}$ are sets of random variables uniformly distributed in $[-1/2,1/2]$, $v>0$ and $w>0$ are the average tunneling amplitudes, and  $d<|v-w|/{\rm max}(v,w)$ is the disorder strength. For $v<w$ the spectrum of the disordered SSH chain still consists of a continuum branch and of two additional discrete levels. The red curve  describes the only occupied discrete level and again is  localized on both edges. In fact, disorder has a stronger impact on the   bulk states, which get  localized too, as shown by the thin black curve in   Fig.\ref{Fig2-chain-groundstate-density-profile}(b), in agreement with Anderson localization\cite{anderson_PR_1958,thouless_1972}. However, their total contribution to the site occupancy profile [blue curve in Fig.\ref{Fig2-chain-groundstate-density-profile}(b)] is still uniformly flat in the bulk and exhibits two dips by the edges, just like in the clean case of Fig.\ref{Fig2-chain-groundstate-density-profile}(a).   
Again, at half filling, the sum of bulk and edge state contributions yields a perfectly uniform occupancy profile $N_{j,s}\equiv 1/2$ [black thick line in Fig.\ref{Fig2-chain-groundstate-density-profile}(b)]. \\
It is also straightforward to understand why such uniform  profile is unaltered by finite temperatures: Despite the energy separation between the two discrete levels is tiny, the  partial occupancy of the level at energy $+\varepsilon^{\rm edge}$ induced by thermal excitations is perfectly balanced by the corresponding depletion of the level at energy $-\varepsilon^{\rm edge}$.

%%%%%%%%%%%%%%%%%%%%%%%%%%%%%%%%%%%%%%%%%% 
%%%%%%%% Local quench from a ring to a chain or viceversa  %%%%%%%%%
%%%%%%%%%%%%%%%%%%%%%%%%%%%%%%%%%%%%%%%%%% 
\section{Breaking charge conjugation in chiral symmetric models: A local quench}
\label{sec-4}
In order to observe some effects of the quench in the real-space occupancy, and possibly the appearance of the topological edge states, a necessary condition is that at least one of the two crucial hypotheses  of the theorem, {Eq.}(\ref{cond-rho}) and {Eq.}(\ref{cond-Ham}), is violated.   Here below we  show how this is possible  while  still {\it preserving the chiral symmetry} and while operating at {\it half-filling}, i.e under the  conditions where  the  SSH model is rigorously characterized as a topological insulator. \\

\subsection{Ring-to-chain quench} 
The first option is to violate the hypothesis of charge-conjugation invariance of the pre-quench state,   {Eq.}(\ref{cond-rho}). This can be achieved by choosing as $\hat{\boldsymbol{\rho}}^{\rm pre}$   a thermal equilibrium state of the homogeneous half-filled SSH model, like in {Eq.}(\ref{rho-pre-thermal}), where $\hat{\mathcal{H}}^{\rm pre}$ is  {Eq.}(\ref{SSH-gen}) defined on a ring-shaped lattice threaded by a magnetic flux $\Phi$. In this case both $\hat{\mathcal{H}}^{\rm pre}$ and $\hat{\boldsymbol{\rho}}^{\rm pre}$ break   time-reversal symmetry~$\mathcal{T}$ and hence  charge conjugation~$\mathcal{C}$ symmetry, so that the condition {Eq.}(\ref{cond-rho}) is violated. Note that, nevertheless, the  pre-quench site occupancy still equals exactly 1/2. Indeed, since $\hat{\mathcal{H}}^{\rm pre}$  
 commutes with the chiral symmetry~$\mathcal{S}$,  the relation
\begin{eqnarray}
N_{j,s}^{\rm pre}  &=& {\rm Tr}\left[ \hat{\boldsymbol{\rho}}^{\rm pre} \hat{n}_{j,s}\right] =\frac{ {\rm Tr}\left[ \mathcal{S} e^{-\beta \hat{\mathcal{H}}^{\rm pre}}  \mathcal{S} ^{-1}  \mathcal{S} \hat{n}_{j,s} \mathcal{S}^{-1} \right]}{ {\rm Tr}\left[  e^{-\beta \hat{\mathcal{H}}^{\rm pre}}   \right]} = \nonumber \\
&=& \frac{{\rm Tr}\left[   e^{-\beta \hat{\mathcal{H}}^{\rm pre}} (1-   \hat{n}_{j,s} ) \right]}{ {\rm Tr}\left[  e^{-\beta \hat{\mathcal{H}}^{\rm pre}}   \right]}    =1-N_{j,s}^{\rm pre} 
\end{eqnarray}
implies that $N_{j,s}^{\rm pre} =1/2 \,\, \,\forall j, s$.  For definiteness, we take for $\hat{\boldsymbol{\rho}}^{\rm pre}$ the ground state of the  SSH ring.

\begin{figure*}[h]
\centering
\includegraphics[width=0.96\linewidth]{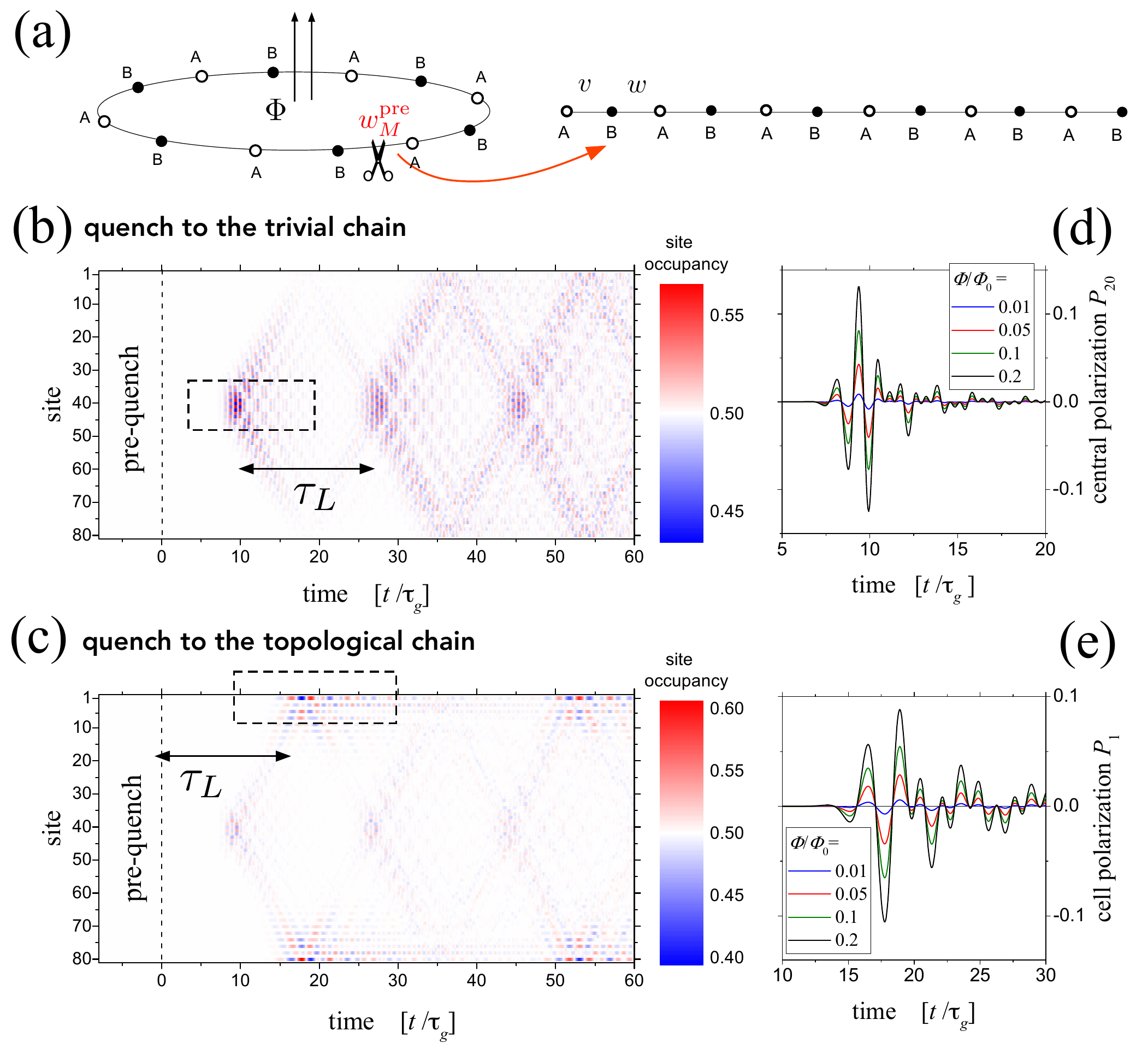}
\caption{\label{Fig3-quench-ring-to-chain} (a) Sketch of a local quench: By cutting a bond $w$ of an SSH ring threaded by  a   flux $\Phi$, a SSH chain is obtained. Here we used a ring with ${N}=80$ sites (i.e. $M=40$ cells). (b) space-time evolution of the site occupancy $N_{j,s}$ for the case $\Phi=\Phi_0/5$ and $w=0.7\,|v|$:   the cut bond $w$ is weak, so that the post-quench  SSH  chain is in the trivial phase; (c) space-time evolution of the site occupancy $N_{j,s}$ for the case $\Phi=\Phi_0/5$ and $|v|=0.7\,|w|$:  the cut bond $w$ is strong, so that the post-quench chain is in the topological phase; (d) the central polarization $P_{20}=N_{20A}-N_{20B}$   is plotted in the time frame highlighted in panel (b) by the dashed box, for various values of the pre-quench ring flux~$\Phi$; (e) the  edge polarization $P_1=N_{1A}-N_{1B}$   is plotted in the time frame highlighted in panel (c) by the dashed box, for various values of  $\Phi$. }
\end{figure*}

Then, after isolating the system from the environment, at $t=0$ we perform a {\it local} quench, i.e. we bring one single ring bond, e.g. $w_M$, from $w_M^{\rm pre}=w$ to $w_M^{\rm post}=0$, leaving all the other bonds $v$ and $w$ unaltered. As a consequence,  the ring gets cut  into a chain, as illustrated in Fig.\ref{Fig3-quench-ring-to-chain}(a).  The post-quench Hamiltonian, being defined on a chain lattice, preserves $\mathcal{T}$ and $\mathcal{C}$ and the second hypothesis {Eq.}(\ref{cond-Ham}) of the theorem is  satisfied. Depending on whether the cut bond is  weak ($|w^{\rm pre}_M|<|v|$) or strong ($|w^{\rm pre}_M|>|v|$), the post quench chain  is in the trivial  or in the topological phase, respectively. For simplicity we shall consider the limit of an instantaneous quench. Still, 
two timescales characterize the post-quench evolution, namely 
\begin{equation}
\tau_g=\frac{\hbar}{||v|-|w ||}\quad,
\end{equation}
which is the timescale related to the inverse  half-gap, and
\begin{equation}\label{tauL}
\tau_L=\frac{\hbar \,M}{{\rm min}(|v |,|w |)}\quad,
\end{equation}
corresponding to the typical time an electron wavepacket takes to travel the system length $L=Ma$ (see the Appendix).

%%%%%%%%%%%%%%%%%%%%%%%
%%%%%%%%%%%%%%%%%%%%%%

The space-time evolution of the   site occupancy induced by the quench is depicted in Fig.\ref{Fig3-quench-ring-to-chain}  for a ring of ${N}=80$ sites ($M=40$ cells), initially threaded by a magnetic flux $\Phi=\Phi_0/5$, where the $w$-bond  between sites 1 and 80 is cut by the local quench. In the plot, the red (blue) color   characterizes a positive (negative) fluctuation $N_{j,s}-1/2$ from the pre-quench occupancy $1/2$ (white color), and time is expressed in units of $\tau_g$. Panels (b) and (c) refer to the cases of quench from the ring to the trivial and to the topological chain, respectively.
After the bond is cut ($t>0$), we observe in both cases  that the site occupancy remains roughly equal to $1/2$ everywhere until a time $\tau_L/2$, which corresponds to the timescale needed by the quench-induced electron waves propagating in opposite directions to meet again and interfere   in the middle of the chain, i.e. at the opposite site of the cut bond. 

After such time,   the two panels  feature    qualitatively different behaviors.  Indeed for a quench to the trivial chain [see Fig.\ref{Fig3-quench-ring-to-chain}(b)], the fluctuations $N_{j,s}-1/2$ from the pre-quench occupancy are more pronounced near the center of the chain and occur at times  $t \simeq \tau_L(m+1/2)$ corresponding to half-integer values  of the typical time related to the chain length, {Eq.}(\ref{tauL}). In contrast,  for a quench to the topological chain [Fig.\ref{Fig3-quench-ring-to-chain}(c)], the largest  occupancy fluctuations  are observed  at the chain edges and dynamically appear for the first time at   $t \simeq \tau_L$, and then again at odd integer multiples $(2m+1)\tau_L$.  Note that, at each appearance, the  fluctuations take  opposite signs at the two boundaries, since the total charge is conserved.

We emphasize that the pre-quench flux $\Phi$ is crucial in determining  the magnitude of fluctuations from the occupancy value $1/2$,  for both quenches to the trivial and to the topological chain. For the quench to the trivial chain this  is clearly illustrated in panel (d), which displays the bulk polarization $P_{20}=N_{20,A}-N_{20,B}$, i.e. the polarization of the   central cell $j=20$, for the time range highlighted by the dashed box of panel (b), for various values of the   flux $\Phi$ through the pre-quench ring. Similarly, for the case of quench to the topological chain, panel (e) shows the edge polarization $P_1=N_{1,A}-N_{1,B}$, i.e. the polarization of the   cell $j=1$ in the time frame highlighted in panel (c). When $\Phi$ is vanishing or equal to  $\Phi=p\Phi_0/2$, with $p\in \mathbb{Z}$, the site occupancy   remains  locked to $N_{j,s}\equiv 1/2$  at any time.

%For  the topological case this is clearly illustrated in panel (d), which displays the edge polarization $P_1=N_{1,A}-N_{1,B}$, i.e. the polarization of the   cell $j=1$, for the time range highlighted by the dashed box of panel (c), for various values of the   flux $\Phi$ through the pre-quench ring.  When $\Phi$ is vanishing or equal to  $\Phi=p\Phi_0/2$, with $p\in \mathbb{Z}$, the site occupancy   remains  locked to $N_{j,s}\equiv 1/2$  at any time.

Furthermore, for a given flux $\Phi$, the magnitude of the fluctuations also depends on the gap $2\varepsilon_g$ through the ratio   $r=\varepsilon_g/{\rm max}(|w|,|v|)$. In particular, when $r \rightarrow 1$, the model tends to the dimerized limit  where the flux plays no role and  the fluctuations from occupancy $1/2$ vanish everywhere. In general, both for quenches to the trivial and to the topological phase, a decrease in the value of $r$ implies  an increase in the fluctuation magnitude. 
For instance, for ${N}=80$ and $\Phi=\Phi_0/5$,   when $r$ is decreased from $r=0.5$ down to $r=0.1$,   the fluctuation  magnitude  at the  edge sites of the topological chain increases from $14\%$ to $30\%$ of the pre-quench site occupancy value $1/2$. 
However, when $r \rightarrow 0$  the model tends to the gapless metallic tight-binding model, and a difference between topological and trivial phase emerges. Indeed if such limit  is taken from the trivial phase ($|v|>|w|$) the magnitude of fluctuations located near the center of the chain [see Fig.\ref{Fig3-quench-ring-to-chain}(b)] tends to a finite value and survive even in the metallic case. In contrast, when the gap is decreased from the topological phase side ($|v|<|w|$), the magnitude of the fluctuations located at the chain edges [see Fig.\ref{Fig3-quench-ring-to-chain}(c)] 
eventually drops to zero for $r<10^{-2}$ and the effect is completely suppressed in the metallic case $r=0$, in agreement with the fact that edge states disappear in such a case. 

%%%%%%%%%%%%%%%%%%%%%%%
\begin{figure}[h]
\centering
\includegraphics[width=0.7\linewidth]{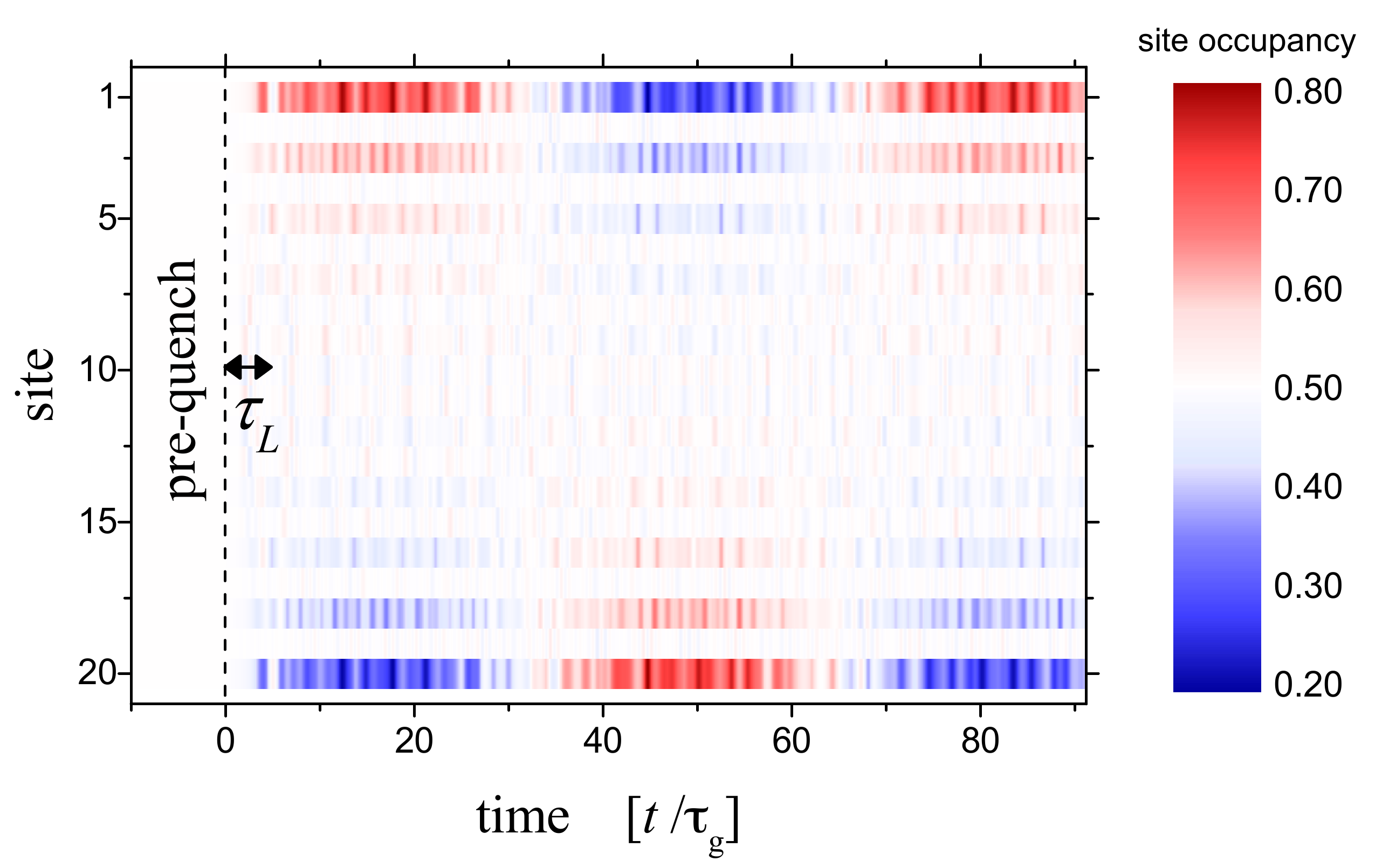}
\caption{\label{Fig4-quench-ring-to-chainN20} Quench from a ring with flux to a chain in the topological phase: The space-time evolution of the site occupancy $N_{j,s}$ in the case of a small site  number (${N}=20$, i.e. $M=10$ cells). For comparison with a longer chain, the parameters $|v|=0.7\,|w|$ and $\Phi=\Phi_0/5$ are the same as in Fig.\ref{Fig3-quench-ring-to-chain}(c).}
\end{figure}
%%%%%%%%%%%%%%%%%%%%%%

To conclude this subsection, we analyze how the  site occupancy fluctuations induced by the quench depend on the  chain length $L=Ma$. The cases analyzed so far in Fig.\ref{Fig3-quench-ring-to-chain} correspond to the regime $\tau_L \gg \tau_g$ of a long chain. In this regime the magnitude of site occupancy fluctuations does not significantly change with the number  ${N}=2M$ of sites, while the  occurrence timescale $\tau_L$    does of course depend on $M$ [see {Eq.}(\ref{tauL})]. For fixed values of $v$ and $w$, when the number of lattice sites is reduced one reaches the regime $\tau_L \sim \tau_g$, where  the typical energy separation $\Delta$ between the bulk states becomes comparable with the gap $2\varepsilon_g$, so that the very notion of bulk gap becomes somewhat questionable. Yet, two discrete energy levels $\pm \varepsilon^{\rm edge}$ near $\varepsilon=0$ are still present in the topological phase.  In  Fig.\ref{Fig4-quench-ring-to-chainN20} we have plotted the space-time evolution of the site occupancy for a short lattice with ${N}=20$ sites (i.e. $M=10$ cells), when  the ring with flux is cut into a   chain in the topological phase, keeping all the other parameters unchanged with respect to the case of Fig.\ref{Fig3-quench-ring-to-chain}(c).
The comparison   shows   two interesting effects of the reduced system size on the dynamics. First, the magnitude of the fluctuations at the edges is bigger in the shorter chain [Fig.\ref{Fig4-quench-ring-to-chainN20}] than in the longer chain   [Fig.\ref{Fig3-quench-ring-to-chain}(c)], highlighting the dynamical alternation of excess and depletion of occupancy at the two edges.   Second,  along with the short  timescale $\tau_L$ determining the roughly periodic occurrence described above, we observe a second longer period that further modulates the fluctuation magnitude. Such a timescale is associated to the small energy splitting $2\varepsilon^{\rm edge}$ between the two discrete edge states. Indeed such energy separation increases  when reducing the system size and, {despite being} much smaller than the gap, it  becomes  visible through this time-dependent modulation.\\

The local quench cutting the ring thus leads  to qualitatively different behaviors in real-space, depending on whether the post-quench chain is in the  trivial or in the topological phase.  In particular, in the quench to a topological chain  the fluctuations of site occupancy are localized at the edges, and alternate in time from excess to depletion. A time-resolved measurement is thus needed to observe such real-space signatures,  while a time-average   would vanish, just like in any site of the bulk. This is typical of a half-filled system.  In Sec.\ref{sec-5} we shall discuss the case of different filling values.

\subsection{Chain-to-ring quench} 
\begin{figure}[h]
\centering
\includegraphics[width=0.8\linewidth]{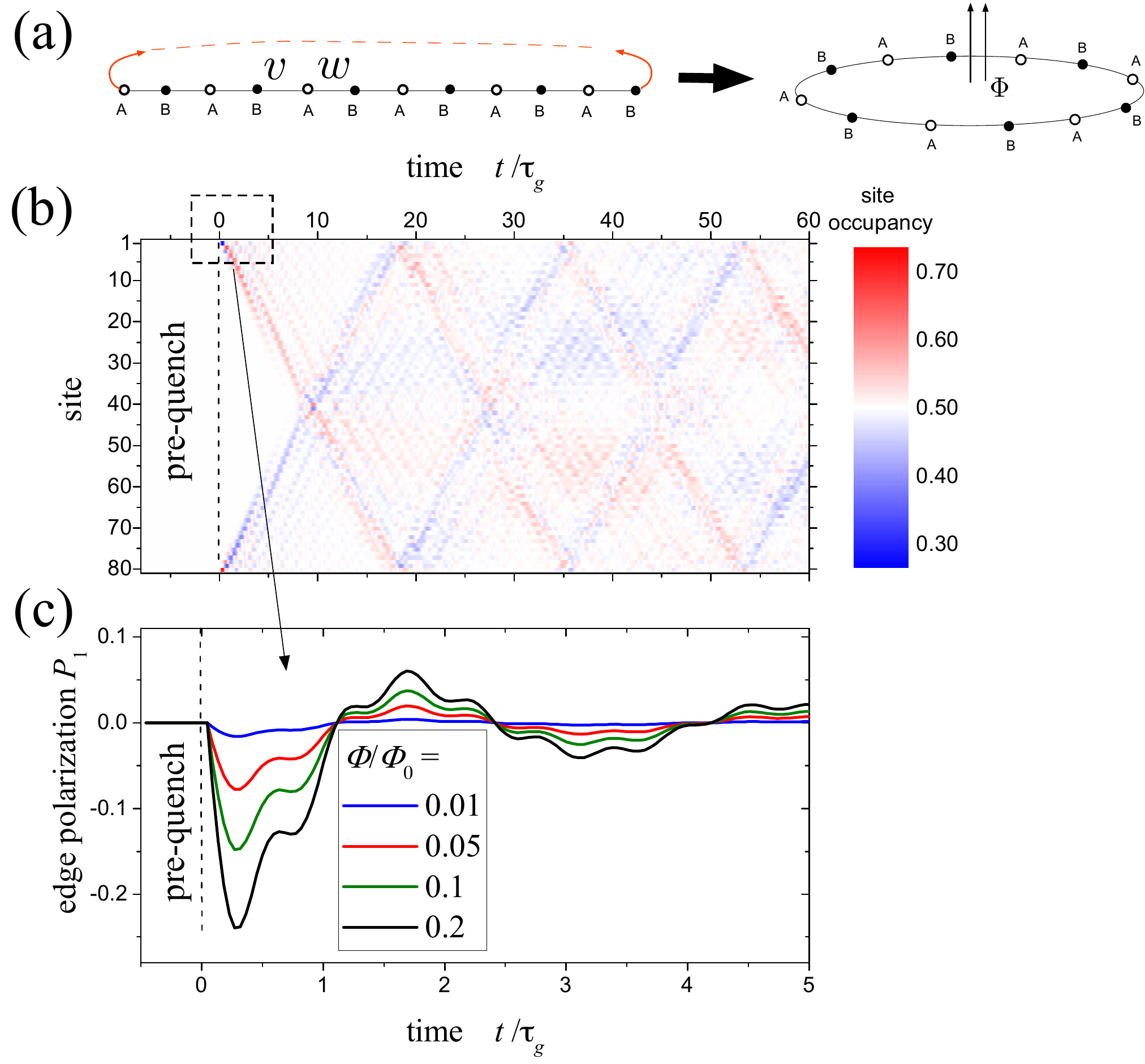}
\caption{\label{Fig5-quench-chain-to-ring} (a) By  binding the extremal sites of a SSH chain  in the topological phase, the chain is brought into   a SSH ring threaded by a  flux. Here we have taken ${N}=80$ sites (i.e. $M=40$ cells) and  $v=0.7\,|w|$; (b)  space-time evolution of the site occupancy $N_{j,s}$ along the ring, for a    flux $\Phi=\Phi_0/5$; (c) time evolution of the edge polarization $P_1=N_{1A}-N_{1B} $ of the first cell ($j=1$), for various values of the   flux   $\Phi$. }
\end{figure}

The second possibility to tackle the theorem of Sec.\ref{sec-theorem} is to break the  hypothesis {Eq.}(\ref{cond-Ham}). 
We keep   the first hypothesis {Eq.}(\ref{cond-rho}) by choosing as a pre-quench state the ground state of a homogeneous half-filled SSH chain. Then, one can  perform a local quench  binding the first and last site of the chain, i.e. bringing the tunneling amplitude $w_M$ from $w^{\rm pre}_M =  0$ to $w^{\rm post}_M =w$, thereby enclosing the chain into a ring, as illustrated in Fig.\ref{Fig5-quench-chain-to-ring}(a). If the ring is threaded by a magnetic flux,  the quenching Hamiltonian breaks  $\mathcal{T}$ and hence $\mathcal{C}$ symmetries,  and the condition {Eq.}(\ref{cond-Ham})   is violated, opening up the possibility to observe real-space signatures of the quench.
%%%%%%%%%%%%%%%%%%%%%%%
%%%%%%%%%%%%%%%%%%%%%%

Figure \ref{Fig5-quench-chain-to-ring}(b) displays the space-time evolution of the site occupancy when the pre-quench state is the ground state of a $80$-sites half-filled SSH chain   in the topological phase ($|v|<|w|$) and the post-quench ring is threaded by a flux $\Phi=\Phi_0/5$. Note that, although the initial chain is in the topological phase, before the quench ($t<0$) the site occupancy is locked to $1/2$ everywhere, in agreement with the theorem proven above, and no signature of the edge state emerges in real-space. However, after the quench ($t>0$), two occupancy fluctuations $N_{j,s}-1/2$ of opposite signs depart from the bridged link and propagate  along the ring in opposite directions, determined by the sign of the flux $\Phi$. 
When the SSH chain is initially in the  trivial phase, such an effect is absent, the occupancy evolution is quite similar  to the case of a ring-to-trivial chain quench already shown in Fig.\ref{Fig3-quench-ring-to-chain}(b) and is not reported here.
The evolution of the edge polarization at the cell $j=1$ is shown in Fig.\ref{Fig5-quench-chain-to-ring}(c) in the early time range highlighted by the dashed circle of panel (b),   for various flux values. Again the presence of the flux   is crucial to observe real-space signatures of the quench.

%%%%%%%%%%%%%%%%%%%%%%%%%%%%%%%%%%%%%%
%%%%%%%%%%%%%%%%%%%%%%%%%%%%%%%%%%%%%%
%%%%%%%%%%%%%%%%%%%%%%%%%%%%%%%%%%%%%%
\section{Quenches in chiral-symmetry broken models and   effects of band filling}
\label{sec-5}
In the previous section we have shown how to violate  the conditions Eqs.(\ref{cond-rho}) and (\ref{cond-Ham}), while preserving the  chiral symmetry~$\mathcal{S}$ in the half-filled SSH model.
Here we want to explore the dynamical effect of quenches when the chiral symmetry is broken and   the filling is not necessarily equal to $1/2$, i.e. beyond the framework where the model can be classified as a topological insulator.
As a matter of fact, in a realistic electron model on a bipartite lattice, 
the chiral symmetry is fragile. A difference $\delta_j$ between the  on-site  energies of   $A$ and $B$ sites is likely to exist, leading to   an additional Hamiltonian term
\begin{equation}
\hat{\mathcal{H}}_{\chi b}\label{H-chi-b}
 = \sum_{j} \delta^{}_j  (\hat{n}^{}_{j,A} -\hat{n}^{}_{j,B})   \quad,
\end{equation}
which breaks   the chiral symmetry  because $\mathcal{S}\hat{\mathcal{H}}_{\chi b}\mathcal{S}^{-1}=-\hat{\mathcal{H}}_{\chi b}$~\cite{nota_Hchib}.  
On the one hand, in the absence of chiral symmetry the very topological classification is not well defined since, for instance, one could go from the range $|v|>|w|$ to the range $|v|<|w|$ without closing the gap. A priori, there is  no guarantee    that the topological states exist at all. On the other hand,   numerical analysis shows that, if the values of the $\delta^{}_j$ are small compared to the band gap $2\varepsilon_g$,  edge states still persist.
Specifically, we shall consider a Hamiltonian
\begin{equation}\label{Htot}
\hat{\mathcal{H}} = 
\hat{\mathcal{H}}_{SSH,\chi}+\hat{\mathcal{H}}_{\chi b} \quad,
\end{equation}
where the first term, {Eq.}(\ref{SSH-gen}), contains a {\it chiral disorder} $w_j=w(1+\xi_j d)$ and $v_j=v(1+\eta_j d)$, while the second term, {Eq.}(\ref{H-chi-b}), contains   a {\it chiral-breaking disorder}  $\delta_j=\zeta_j \,{\rm max}(v,w)d$. Here we have assumed $v,w>0$ and  $\{\xi_j , \eta_j, \zeta_j\}$ denote real random variables   uniformly distributed in $[-1/2,+1/2]$ with a disorder strength $d<|v-w|/{\rm max}(v,w)$. 
Note that, although for each disorder realization the chiral symmetry is   broken by {Eq.}(\ref{Htot}), the disorder-averaged Hamiltonian {Eq.}(\ref{Htot}) still preserves the chiral symmetry, so that  the trivial and topological phases  can still be defined in the sense of the average values $v$ and $w$. 
While Ref.\cite{chien_2016} analyzed the Hamiltonian {Eq.}(\ref{Htot}) in the case of local quenches performed over a  long quench time  ($\tau_q \gg \tau_g$) and at half filling, here we shall focus on the complementary situation of a global quench in the short quench time limit ($\tau_q \ll \tau_g$) and consider also filling values different from  a half, which turns out to be important for the effects in real-space.

Let us thus go back to the original problem   illustrated in Fig.\ref{Fig1-chain-quench-scheme}, and analyze a global quench from the trivial to the topological   chain,   where now the  Hamiltonian  {Eq.}(\ref{Htot})  includes  the chiral-breaking term  {Eq.}(\ref{H-chi-b}).\\

%%%%%%%%%%%%%%%%%%%%%%%
\begin{figure}[h]
\centering
\includegraphics[width=0.6\linewidth]{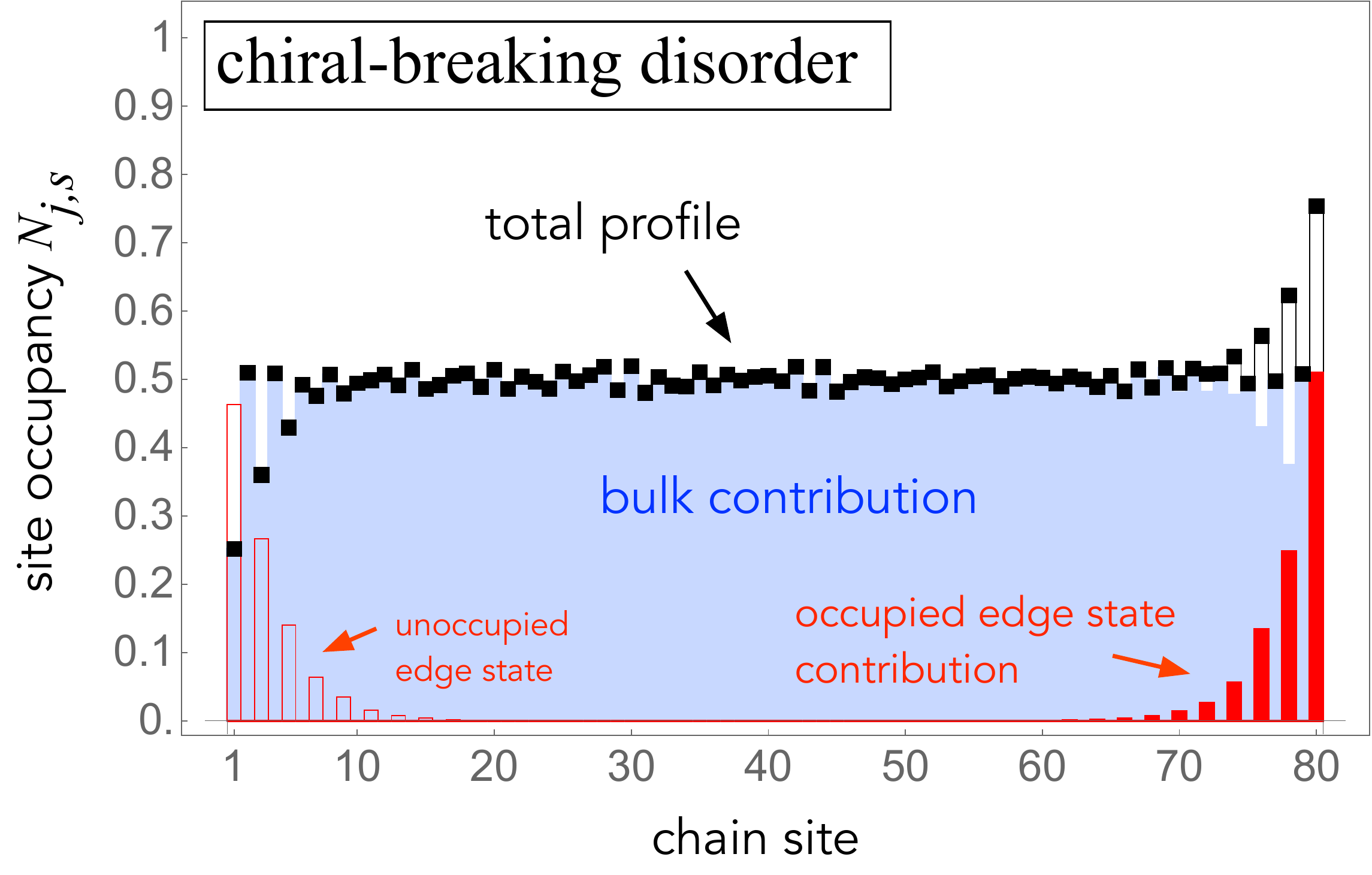}
\caption{\label{Fig6-edge-wavefunctions} The  site occupancy profile  of the half-filled chain of  model (\ref{Htot}) in the topological phase. The parameters are the same as in Fig.\ref{Fig2-chain-groundstate-density-profile}, with the addition of the chiral-breaking disorder term. Its effect is to localize the two edge state wavefunctions   on opposite edges (thick and thin red curves), differently from the case with purely chiral disorder [see Fig.\ref{Fig2-chain-groundstate-density-profile}(b)]. The contribution of all occupied bulk states is described by the blue curve. The total site occupancy profile, depicted by the   thick solid black curve, exhibits a peak at one edge and a depletion at the other edge.  }
\end{figure}

%%%%%%%%%%%%%%%%%%%%%%

\subsection{The case of half-filling}
At first, one might  even expect  that $\hat{\mathcal{H}}_{\chi b}$  may   favor the  appearance of the edge states already at half-filling. Indeed without such term {Eq.}(\ref{H-chi-b})    the site occupancy would  always  remain strictly locked to $1/2$, due to the  theorem proven in Sec.\ref{sec-theorem}.  In contrast,  because the  term (\ref{H-chi-b})
also  breaks the charge conjugation symmetry,  $\mathcal{C}\hat{\mathcal{H}}_{\chi b}\mathcal{C}^{-1}=-\hat{\mathcal{H}}_{\chi b}$, 
 the hypotheses of the theorem are violated, opening up  the possibility to observe fluctuations of the site occupancy, possibly at the edges.
This expectation seems to be confirmed when analyzing how the  edge state  wavefunctions are modified by the term  $\hat{\mathcal{H}}_{\chi b}$. While in the purely chiral SSH model $\hat{\mathcal{H}}_{SSH,\chi}$  each wavefunction is  localized on {\it both} edges even in the presence of {\it chiral disorder} $\{ v_j,w_j\}$  [see Fig.\ref{Fig2-chain-groundstate-density-profile}], the addition  of the    chiral-breaking disorder $\hat{\mathcal{H}}_{\chi b}$ localizes   each discrete state of the   Hamiltonian (\ref{Htot})   only on {\it one single} edge of the chain, as shown by the red curves of Fig.\ref{Fig6-edge-wavefunctions}. This is the hallmark of the break-up of the topological protection.  Depending on the specific disorder realization, one of the two edge wavefunctions is energetically slightly more favoured than the other, so that the ground state of the half-filled topological chain with chiral-breaking disorder   exhibits a site occupancy with an enhancement at one edge, a depletion on the other edge and a value roughly equal to $1/2$ in the bulk [see black curve of Fig.\ref{Fig6-edge-wavefunctions}]. 
In principle, such real-space signature of the edge state should appear by performing an infinitely slow quench from the  trivial  to the topological chain, where the pre-quench trivial ground state should evolve into the post-quench topological ground state. However, if the chiral-breaking disorder term is weak, the energy separation between the two localized states is very small. In practice, at half filling, any finite temperature in the pre-quench state leads the post-quench state to exhibit only {\it half-occupancy} of {\it both} discrete states, quite similarly to what happens in the   chiral SSH model $\hat{\mathcal{H}}_{\chi b}$.

%%%%%%%%%%%%%%%%%%%%%%%
\begin{figure*}[h]
\centering
\includegraphics[width=.95\linewidth]{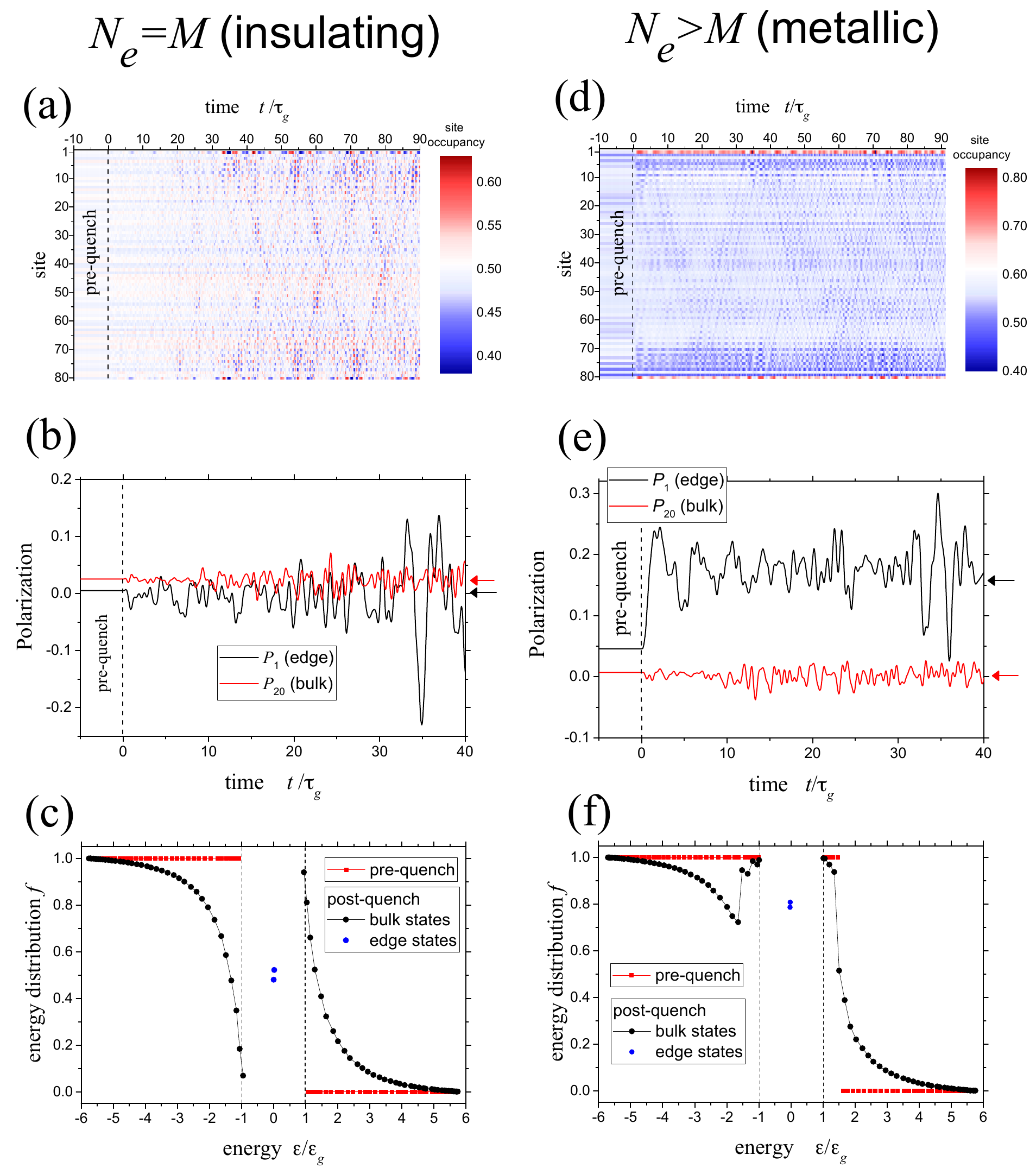}
\caption{\label{Fig7-disorder} Effects of an instantaneous quench in a 80-sites chain ($M=40$ cells) from the  trivial phase ($w^{\rm pre}=0.7 v^{\rm pre}$) to the topological phase ($w^{\rm post}=v^{\rm pre}$ and $v^{\rm post}=w^{\rm pre}$), for a given realization of both chiral and chiral-breaking disorder with strength $d=0.1$. Panels (a)-(b)-(c) refer to the case $N_e=M$ ($\mu=0$, half filling,  insulating pre-quench state), while panels (d)-(e)-(f) to the case $N_e>M$  ($\mu= 0.44 \,v^{\rm pre}$, metallic pre-quench state).  Panels (a) and (d)   describe the space-time evolution of the site occupancy. Panels (b) and (e) display the polarizations $P_1$ and $P_{20}$ of the edge cell ($j=1$) and the central cell ($j=20$). Panels (c) and (f) show the nonequilibrium distribution of the post-quench Hamiltonian (black and blue circles for bulk and edge states, respectively). For comparison, the pre-quench equilibrium distribution is shown in red squares.}
\end{figure*}
%%%%%%%%%%%%%%%%%%%%%%

{The same effect occurs when the duration of the quench is short, as} shown in Fig.\ref{Fig7-disorder}(a), which  illustrates the space-time evolution of the site occupancy of  a 80-sites chain that is suddenly quenched from the {ground state of the} trivial phase ($w^{\rm pre}=0.7 v^{\rm pre}$) to the topological phase   ($w^{\rm post}=v^{\rm pre}$ and $v^{\rm post}=w^{\rm pre}$), for given pre-quench and post-quench realizations of chiral and chiral-breaking disorder with strength $d=0.1$. Due to the chiral-breaking term {Eq.}(\ref{H-chi-b}), static deviations from the site occupancy $1/2$ are present even before the quench, while after the quench  these deviations fluctuate in time as well.
In Fig.\ref{Fig7-disorder}(b),  the corresponding  edge polarization $P_1=N_{1,A}-N_{1,B}$ (black curve) is compared to the polarization at the central chain cell $P_{20}=N_{20,A}-N_{20, B}$ (red curve). As one can see,   fluctuations  do have  larger amplitudes at the chain edges than in the chain bulk. However, at each edge, the  site occupancy experiences an alternation of   depletion and excess [blue and red colors in panel (a)],  just like for the quenches preserving the chiral symmetry discussed above  [see e.g. Fig.\ref{Fig3-quench-ring-to-chain}(b) or Fig.\ref{Fig4-quench-ring-to-chainN20}].   This quantitatively shows that   the chiral-breaking term does not really improve the  observability of real-space effects of the quench. In particular, the time-average of the fluctuations at the edge and in the bulk    is essentially the same, as highlighted by the  arrows in Fig.\ref{Fig7-disorder}(b).   

In  energy-space, however, the effects of the quench are seizable. Indeed the  energy distribution of the post-quench Hamiltonian displayed   in Fig.\ref{Fig7-disorder}(c) strongly differs from the pre-quench equilibrium one, depicted in red for comparison. In particular, as far as the  continuum spectrum is concerned (black symbols), the  non-equilibrium distribution that we obtain for the disordered chain-to-chain quench is quite similar to the result obtained for a quench in the clean  bulk SSH model analyzed in Ref.\cite{sassetti_PRB2018}, and describes the population inversion effect found upon   quenching from one phase to the other. This effect can open up the possibility that, when the SSH model is coupled to a radiation\cite{gebhard}, a stimulated  emission occurs due to transitions from the (almost filled) continuum states near the bottom of the conduction band to the (almost depleted) states near the top of the valence band, with a radiation frequency corresponding to the band gap. Apart from the presence of disorder, the major difference from the ring case arises from the presence of the topological edge states in the post-quench spectrum of the chain, highlighted by the blue symbols. Note that their occupancy is roughly $1/2$. On the one hand, this is precisely what disguises these states   in  the  real-space occupancy at the chain edges, as argued above [see Fig.\ref{Fig7-disorder}(a)]. On the other hand,   differently from a purely bulk SSH system, in a chain quenched to the topological phase the presence of half-occupied discrete levels  near $\varepsilon=0$ causes  an additional emission process, characterized by a frequency corresponding to a half of the gap, similar to the phenomenon described in Ref.\cite{TO-GE}  of a quenched quantum well potential.

\subsection{Away from half-filling}
The results obtained above at half filling ($N_e = M$), where the  model (\ref{Htot}) describes a band insulator,  indicate that a quench from a trivial to the topological phase of a chain  does lead to the appearance of  site occupancy fluctuations that are larger at the chain edges than in the bulk. However, the alternation of    depletion  and excess yields a vanishing result upon time-average, both at the edges and  in the bulk.  
In the short quench time limit  this holds  for any temperature of the pre-quench state. 
We now analyze the effects of a quench in a non half-filled chain, where the number $N_e$ of electrons differs from  the number $M$ of lattice cells. Note that in such a case the pre-quench ground state is  {\it metallic}, with an excess of electrons (holes) in the conduction (valence) band for $N_e>M$ ($N_e<M$). At finite temperature this is described by
\begin{equation} 
\hat{\boldsymbol{\rho}}^{\rm pre} = \frac{e^{-\beta (\hat{\mathcal{H}}^{\rm pre}-\mu \hat{N}_e)}}{{\rm Tr}[e^{-\beta(\hat{\mathcal{H}}^{\rm pre}-\mu \hat{N}_e)}]} \quad,
\end{equation}
where  $\hat{N}_e=\sum_{j,s} \hat{n}_{j,s}$ is the total electron number operator and $\mu$ the chemical potential. While in the insulator $\mu=0$, the metallic state is described by $\mu \neq 0$. Moreover, because $\hat{N}_e$ transforms under charge conjugation as $\mathcal{C} \hat{N}_e \mathcal{C}^{-1} =2 M - \hat{N}_e$    the condition (\ref{cond-rho}) of the theorem proven in Sec.\ref{sec-theorem} is violated, opening up the way to observe effects of the quench in real-space.

In principle, if one takes the standard disorder-free SSH model {Eq.}(\ref{SSH-clean}), the ideal situation to observe the appearance of the topological edge states is a Gedankenexperiment where the pre-quench state is the ground state of the trivial SSH chain ($v^{\rm pre}>w^{\rm pre}$) with exactly $N_e= M+1$ electrons, with the extra electron lying in the  conduction band. By performing  an infinitely slow quench ($\tau_q \rightarrow \infty$) to the topological chain ($w^{\rm post}>v^{\rm post}$), the ground state  evolves to the post-quench ground state of the topological chain [see Fig.\ref{Fig1-chain-quench-scheme}], where now {\it both} discrete levels will be occupied, instead of only one like in the half-filled case. The extra electronic level, delocalized on both edges, causes the gradual appearance of peaks localized at the chain edges in the site occupancy profile, over a value of $N_{j,s}=1/2$ in the bulk of the chain.
In practice, however, such ideal conditions are  not necessarily easy to realize and/or useful. First, chiral-breaking disorder {Eq.}(\ref{H-chi-b}) is typically present as well. Second, in view of technological applications, one typically wants these operations to be performed sufficiently fast. For a finite  and possibly short quench time $\tau_q$   the post-quench state may differ from the slow-quench scenario. Third,   in a metallic system the filling would deviate from $1/2$ not just by one single electron. A finite fraction of the (say) conduction band is occupied and, even in the adiabatic quench limit,
these extra conduction states may mask the localized peaks due to the edge states. This is certainly the case, for instance, when the filling approaches 1.  
The question is thus whether   the edge states dynamically appear in real-space when these  aspects are taken into account.

For definiteness,  we shall analyze the case $N_e>M$, where before the quench a small fraction of the conduction band of the trivial phase  is occupied, e.g. $1/10$ of the conduction bandwidth from its band bottom. For the values $w^{\rm pre}=0.7 v^{\rm pre}$ this corresponds to setting the chemical potential to $\mu=0.44 \,v^{\rm pre}$. Then, we consider a quench   to the topological phase of the chain ($w^{\rm post/pre}=v^{\rm pre/post}$), in the short quench time limit ($\tau_q\rightarrow 0$). The resulting time evolution of the site occupancy is shown in Fig.\ref{Fig7-disorder}(d). As one can see, while in the pre-quench state the edge states are absent, after the quench they start to become visible and stable.  This can be seen explicitly in Fig.\ref{Fig7-disorder}(e), where the polarization of the edge cell (black curve) and of the central chain cell (red curve) are compared. Differently from the half-filling case [panel(b)], the edge  occupancy oscillates around an average value that is finite and thus differs from the small one obtained in the bulk of the chain, as highlighted by the arrows.  Finally Fig.\ref{Fig7-disorder}(f) shows the corresponding nonequilibrium post-quench energy distribution. Note that the occupancy of the discrete states near $\varepsilon=0$ significantly differs from $1/2$. On the one hand, this is the reason for their appearance in real-space at the chain edges. On the other hand, this reduces the occupancy difference from the (almost filled) states near the bottom of the conduction band, thereby reducing the spontaneous emission effect as compared to the half-filling case shown panel (c).\\

\section{Discussion and conclusion}
\label{sec-6}
In this article we have analyzed how a quantum quench applied to  the SSH model (\ref{SSH-gen}) impacts  on  observables that are local in real-space, namely the site occupancy and the cell polarization.  

In Sec.\ref{sec-3} we have proven a general theorem ensuring that, when the pre-quench state and the quenching Hamiltonian fulfill the charge conjugation symmetry $\mathcal{C}$,  the occupancy of each lattice site  remains firmly locked to the value $1/2$, at any time. 
These symmetries are always satisfied in the customary case of a half-filled SSH chain. Indeed, because $\mathcal{C}=\mathcal{S}\mathcal{T}$ and the chiral symmetry $\mathcal{S}$ is preserved by the SSH model, the breaking of $\mathcal{C}$ requires also the breaking of time-reversal symmetry $\mathcal{T}$, which is not possible for spinless electrons in a chain lattice with OBCs.
As a  consequence of the proven theorem,  a  quench from the trivial to the topological phase performed on a SSH chain has no effect whatsoever {on the site occupancies.    
In particular, no signature of  the topological edge states appears locally in real-space, independently of the quench protocol, of the temperature of the pre-quench thermal state and also of the presence of chiral disorder. This is   strikingly different from what is known to happen in $k$-space. Indeed when a quench between two topologically different phases is performed,  a dynamical quantum phase transition\cite{Heyl_2013} is known to arise, and the momentum distribution exhibits a band population inversion related to a dynamical topological invariant\cite{Vajna_2015,sassetti_PRB2018,Yang_2018}.   Our result thus implies that these out of equilibrium phenomena can be detected in real space only through   {\it non-local} quantities, such as correlation functions. }

{The  effects of the quench can become visible in  observables that are local in real space only when charge conjugation symmetry is broken. }  
This can be done either remaining within the framework of the topological insulator characterization, i.e. by preserving the chiral symmetry $\mathcal{S}$   and the half-filling condition, or going out of such framework. 
The first case requires    suitably engineered setups. 
   In particular, in Sec.\ref{sec-4}  we have  shown that 
  a {\it local} quench cutting a SSH ring threaded by a flux into a SSH chain (or viceversa), violates the hypotheses of the above theorem while still remaining in the topological insulator framework. Real-space effects of the quench then become   observable and quite distinct dynamical features appear in the two phases. In particular, when   the pre-quench ring is cut into a trivial   chain the site occupancy fluctuations  appear near the center of the chain, while when the ring is cut into a topological chain these fluctuations appear at the chain edges after a time $\tau_L$
   [see Fig.\ref{Fig3-quench-ring-to-chain}], and then repeat  with a dynamical alternation of excess and depletion at each edge. Such effect at the edges appears even more clearly in a lattice with smaller number of sites [see Fig.\ref{Fig4-quench-ring-to-chainN20}]. Conversely,   when the local quench   bridges a chain   to form a ring with flux, the  site occupancy fluctuations propagate towards the chain center if the pre-quench chain is in the  topological phase[see Fig.\ref{Fig5-quench-chain-to-ring}], whereas   such effect is absent if the chain  is in the trivial phase.    In all such dynamical effects the presence of the flux $\Phi$ threading the ring is crucial. For vanishing flux or for $\Phi=p\Phi_0/2$, where time-reversal and charge conjugation hold, the site occupancy remains locked to the pre-quench value~$1/2$.

In Sec.\ref{sec-5} we have explored the effects of the quench beyond the framework of the topological insulator. By adding a disordered on-site potential term {Eq.}(\ref{H-chi-b})  both the chiral and the charge conjugation symmetries get broken. Although for a disorder realization the topological classification is in principle not well defined and the existence of the edge states is not guaranteed, the disorder-averaged Hamiltonian (\ref{Htot}) still preserves $\mathcal{S}$, and the topological phases can still be considered to hold for weak enough disorder. We have thus analyzed the effects of  a quench from a trivial to a topological chain.
 Our results show that, although the chiral-breaking disorder localizes each edge state wavefunction on one edge only [see Fig.\ref{Fig6-edge-wavefunctions}], in practice such term does not lead to any improvement in terms of their observability in real-space as compared to the purely chiral SSH model. 
In particular, for a half-filled system,  while the  edge   occupancy exhibits much larger fluctuations than  the bulk, its time average is roughly equal to the bulk one [see Fig.\ref{Fig7-disorder}(b)]. Thus,   real-space effects of the quench do exist, but time-resolved measurements are  needed to probe the quench-induced appearance of the edge states.
In contrast,  for filling values different from $1/2$, where the model is slightly metallic, the edge site occupancy fluctuates around a value that is different from the   bulk.
In this case  the dynamical signature of the topological states survive both time-average and  the presence of chiral-breaking disorder,  and persist even in the short quench time limit  [see Fig.\ref{Fig7-disorder}(d)-(e)].  
 
In conclusion our analysis points out that, when a topological insulator is driven out of equilibrium by a quantum quench, the presence of additional symmetries {(such as charge conjugation or time-reversal)} in the quenching Hamiltonian and in the pre-quench state can completely mask the impact of the quench in real-space {occupancies}, even in   customary cases where the energy distributions are typically strongly affected. {Only when such additional symmetries are suitably broken, like} in the setups and protocols proposed {here in Secs. \ref{sec-4} and \ref{sec-5}}, real-space effects do emerge  {in local observables,} and exhibit distinct dynamical behavior in the topological and trivial phases. 
The huge advances in realizing topological models with cold atoms in optical lattices, which nowadays also enable one to effectively implement a Peierls substitution in tunneling amplitudes\cite{spielman_2012,lewenstein_2012,paredes_2017}, represent a promising perspective to test the predicted quench effects.

%%%%%%%%%%%%%%%%%%%%%%%%
%%%%%           APPENDIX         %%%%%%%
%%%%%%%%%%%%%%%%%%%%%%%%
\appendix 

\section*{Appendix} 
\setcounter{section}{1}
\label{appendix}
Here we briefly summarize some aspects of the standard {\it homogeneous} SSH model, which corresponds to taking $v_j\equiv v \in \mathbb{C}$ and $w_j\equiv w\in \mathbb{C}$ in {Eq.}(\ref{SSH-gen}).   Let us now recall the cases of a ring and of a chain.

\subsection {Ring (periodic boundary conditions) and topological classification}
Let    $M$ denote the number of cells and by ${N}=2 M$ the number of lattice sites of   a ring-shaped lattice.   By re-expressing the site operators as $c_{j,s}=M^{-1/2}\sum_k e^{i k j a} c_{k,s}$,  the periodic boundary condition   $c_{M+1,A}=c_{1,A}$  enables one to  straightforwardly  rewrite the Hamiltonian {Eq.}(\ref{SSH-gen}) as a decoupled set of   $k$-dependent Hamiltonians
\begin{equation}\label{SSH-k-clean}
\hat{\mathcal{H}}_{SSH,\chi}=\sum_{k}  \left( \hat{c}^{\dagger}_{k,A}, \hat{c}^{\dagger}_{k,B} \right) H(k) \left(
\begin{array}{c}
c^{ }_{k,A} \\ c^{ }_{k,B}
\end{array} 
\right)   \quad.
\end{equation}
Here $a$ denotes the  cell size,  $k=2\pi n/M a$ the wavevectors (with $n=-\frac{M}{2}, -\frac{M}{2} +1,\ldots, \frac{M}{2}-1$ for even $M$ and $n=-[\frac{M}{2}], -[\frac{M}{2}] +1,\ldots,+[\frac{M}{2}]$ for odd $M$), and $c_{k,s}$ are the Fourier mode operators~\cite{nota-su-k-mode-transformation}. In {Eq.}(\ref{SSH-k-clean})
  $H(k)= \boldsymbol\sigma \cdot \mathbf{b}(k)$  is   the first-quantized SSH Hamiltonian in $k$-space, with $\boldsymbol\sigma $ denoting the set of Pauli matrices and 
$
  \mathbf{b}(k)=\left(   {\rm Re}(v)+ {\rm Re}(w e^{i ka}) \,  ,\,  - {\rm Im}(v)+{\rm Im}(w e^{i ka}) \,, 0  \right)
$ 
 a vector lying in the plane. 
The absence of the $b_z$ component   is the hallmark of the chiral symmetry, which is expressed by the property $\sigma_z H(k) \sigma_z =-H(k)$ in terms of the first-quantized Hamiltonian. 
In terms of the tenfold classification scheme\cite{ludwig-schnyder-ryu_NJP,schnyder-ryu_RMP}, the SSH model is in the AIII symmetry class. If $w,v \in \mathbb{R}$ the Hamiltonian $H(k)$ also fulfills  the properties ${H}^{*}(k)  ={H}(-k)$ and $\sigma_z {H}^{*}(k) \sigma_z=-{H}(-k)$ encoding the time-reversal and charge conjugation symmetries, respectively, and the model is  in the BDI symmetry class.
The spectrum consists of two bands 
\begin{eqnarray}
\varepsilon_\pm(k)&=&\pm|\mathbf{b}(k)|= \\
&=&\pm\sqrt{|w|^2+|v|^2+2 |v w|\cos(k a+{\rm arg}(v)+{\rm arg}(w))}\quad,  \nonumber
\end{eqnarray}
separated by the bandgap $2\varepsilon_g=2||v|-|w||$. The maximal value of the group velocity ${\rm v}(k)=\hbar^{-1}\partial \varepsilon/\partial k$ is ${\rm v}_{max}= a\, {\rm min}(|v|,|w|)/\hbar$  
and determines the minimal timescale {Eq.}(\ref{tauL})
an electron wavepacket takes to travel across the ring length $L=Ma$.  

The  single-particle eigenvectors related to the two bands $\varepsilon_\pm$ are $|u_{-}(k)\rangle =(1\,,\, -e^{i \varphi(k)})^T/\sqrt{2}$ and $|u_{+}(k)\rangle =(e^{-i \varphi(k)}\,,\,1 )^T/\sqrt{2}$, with $\varphi(k)=-\varphi(-k)$ denoting the polar angle of the $\mathbf{b}(k)$ vector, so that $\tan\varphi(k)=b_y(k)/b_x(k)$.
In particular, in  the thermodynamic limit, where $k$ becomes a continuous variable spanning the Brillouin zone $[-\pi,+\pi]/a$, $\mathbf{b}(k)$ draws a circle centered at $({\rm Re}(v),-{\rm Im}(v))$ and with radius $|w|$. As is well known, this enables one to identify two topological  classes of the insulator, depending on whether such circle encloses or not the origin (corresponding to the gap closing). Correspondingly,  the winding number of the fully occupied lower band of the insulator 
\begin{equation}
\nu=-\frac{i}{\pi} \oint \langle u_{-} |\partial_k u_{-}\rangle\, dk=\frac{1}{2\pi} \oint \frac{d\varphi}{dk} \, dk 
\end{equation}
takes two different integer values $\nu=1$ (for  $|v|<|w|$) and $\nu=0$ (for    $|v|>|w|$).
We emphasize that, while the two phases are topologically distinct,  labelling one phase as ``topological" and the other one as ``trivial" is in fact unphysical  as long as the model is defined on a ring. This is because  the topological classification is defined once the   unit cell is identified, which is completely arbitrary in a system with  PBCs, though. Indeed the very  Hamiltonian $\hat{\mathcal{H}}_{SSH}$, written in   {Eq.}(\ref{SSH-clean}) by adopting  $(A,B)$ as  unit cell,   could be equivalently rewritten  choosing   $(B,A)$ as  unit cell, which  would amount  to exchanging the role of intra- and inter-cell hopping amplitudes ($v\leftrightarrow w$), so that the  ``topological" phase for the choice $(A,B)$ corresponds to the ``trivial" phase for the   choice $(B,A)$ and viceversa. 
The emergence of topological edge states, which are perhaps the most striking hallmark  distinguishing the topological character of the two phases, requires the breaking of the PBCs.

\subsection{Chain (open boundary conditions)}
The customary way to break the PBCs is to cut  the ring into a finite chain, thereby interfacing the   SSH model with vacuum. In turn, this lifts the degeneracy about the choice of the unit cell: If (say)  $(A,B)$ is the unit cell of the chain with an even number of sites ${N}=2M$, the OBCs of the chain   impose $c_{M+1,A}=0=c_{0,B}$, where  $M$ again denotes the number of cells.  As argued above, in a chain the hopping amplitudes can be taken as real and positive, $v,w \in \mathbb{R}^+$, without loss of generality. 
The OBCs modify the spectrum and enable one to identify   the actual  topological and trivial phases. Indeed, besides a continuum spectrum similar to the ring, when $v<w$  the chain   also features two additional discrete levels (topological phase), which  are absent for  $v>w$ (trivial phase) instead.
The difference between the two phases becomes apparent in the so called dimerized limit, where one of the two hopping amplitude is set to zero.
The  SSH chain  eigenstates resulting from the OBCc are non-degenerate and can be given an analytic expression\cite{montambaux_2011,hilke-mackenzie_pla_2021}. In particular, the continuum eigenstates  extend  over the entire bulk of the chain and can formally be built by  linearly combining the $|u_{\pm}(k)\rangle$ and $|u_{\pm}(-k)\rangle$ of the ring
\begin{equation}\label{eigenvec-bulk-chain}
|v^{bulk}_{\pm}(k)\rangle= \frac{1}{\sqrt{\mathcal{N}_k}} \sum_{j=1}^M \left( \begin{array}{c} \sin[k a j -\varphi(k)] \\ \pm \sin[k a j] \end{array}\right) 
\end{equation}
where $\tan\varphi(k)=w \sin( ka)/(v+w \cos(k a))$ and $\mathcal{N}_k=M+v (v+w \cos(ka ))/(v^2+w^2+2 v w \cos(k a))$ is a a normalization constant. However, the quantization rule of $k$'s   differs from the customary $k a M=2\pi n$ in the ring and fulfill the transcendental equation $k a (M+1)=\pi n +\varphi(k)$ with $n=1, 2 \ldots M$.   
In contrast, the topological edge eigenstates read\cite{montambaux_2011,hilke-mackenzie_pla_2021}
\begin{equation}\label{eigenvec-edge-chain}
|\psi^{edge}_{\pm}\rangle= \frac{1}{\sqrt{\mathcal{N}_0}}\sum_{j=1}^M (-1)^{j+1} \left( \begin{array}{c}   \sinh[\kappa a (M+1-j)]   \\ \pm \sinh[\kappa a j ] \end{array}\right) 
\end{equation}
where  $\kappa$ fulfills $v \sinh[\kappa (M+1) a]=w\sinh[\kappa M a]$ and $\mathcal{N}_0=(w \sinh(2 \kappa a M)/2v \sinh(\kappa a))-(M+1)$ is a normalization constant.  
They are localized mainly on $A$ sites on the left edge and on $B$ sites on the right edge. Their energies are $\varepsilon^{\rm edge}_{\pm}=\pm  w  \sinh(\kappa a)/ \sinh(\kappa (M+1) a)$, whose difference decreases exponentially as $\sim \exp[-\kappa M a]$ with the number of cells.\\


\section*{References}
\begin{thebibliography}{4}
%%%%%%%%%%%%%%%%%%%
%%%% reviews about topological insulators %%%%%
%%%%%%%%%%%%%%%%%%%%%%%%%%%
\bibitem{hasan-kane_review} Hasan M Z, and Kane C L, 2010 Rev. Mod. Phys. {\bf 82} 3045 
\bibitem{zhang_review}  Qi X-L, and Zhang S-C 2011 Rev. Mod. Phys. {\bf 83} 1057
\bibitem{ando_review} Ando Y 2013 J. Phys. Soc. Jpn {\bf 82} 102001
\bibitem{hasan-moore_review} Hasan  M Z, and Moore J E 2011  Ann. Rev. Cond. Mat. Phys. {\bf  2} 55 
%%%%% reviews about Majoranas  %%%%%%%%
\bibitem{alicea_review} Alicea J 2012 Rep. Progr. Phys {\bf 75} 076501
\bibitem{aguado_review} Aguado R 2017 Riv. Nuovo Cim. {\bf 40} 523

%%%%% experimental observation of Majoranas in nanowires %%%%%%%%
\bibitem{kouwenhoven_2012}   Mourik V, Zuo K,  Frolov S M,  Plissard S R,   Bakkers E P A M, and  Kouwenhoven L P 2012 Science {\bf 336} 1003 
\bibitem{kouwenhoven_2018}  G\"ul \"O,      Zhang H,    Bommer J D S,  de Moor M W A,   Car D,   Plissard  S R,  Bakkers E  P A M,  Geresdi A,  Watanabe K,  Taniguchi T,  and   Kouwenhoven L P 1018 Nat. Nanotech. {\bf 13} 192
\bibitem{frolov_2021}  Yu   P,   Chen J,  Gomanko M,  Badawy G,   Bakkers E P A M,  Zuo K, Mourik V,  and  Frolov S M 2021, Nat. Phys. {\bf 17} 482




%%%%%%%%%%%%%%%%%%%%%%%%%%%%%%%%
%%%% real-space imaging of topological edge states %%%%%
%%%%%%%%%%%%%%%%%%%%%%%%%%%%%%%%
%%%% for Majoranas %%%%%%
\bibitem{yazdani_2014}  Nadj-Perge S, Drozdov I K, Li J, Chen H,   Jeon S,   Seo J, MacDonald A H,  Bernevig A, and  Yazdani A 2014 Science {\bf 347} 602
%%%% other ones %%%%%%
\bibitem{molenkamp_PRX_2013} K\"onig M, Baenninger M, Garcia A G F, Harjee N, Pruitt B L, Ames C, Leubner Ph., Br\"une Ch,  Buhmann H,
Molenkamp L W, and Goldhaber-Gordon D 2013 Phys. Rev. X {\bf 3} 021003
\bibitem{molenkamp_naturemat_2013}  Nowack K C,   Spanton E M,   Baenninger M, K\"onig M, Kirtley J R, Kalisky B,  Ames C,  Leubner Ph., Br\"une Ch,  Buhmann H, Molenkamp L W, Goldhaber-Gordon D, and Moler K A 2013 Nat. Mater.  {\bf 12}, 787
\bibitem{goldman_2013} Goldman N, Dalibard  J, Dauphin  A, Gerbier F, Lewenstein  M,  Zoller  P, and Spielman  I B 2013 Proc. Nat. Acad. Sci. {\bf 110}  6736
%\bibitem{yazdani_2014}  Nadj-Perge S, Drozdov I K, Li J, Chen H,   Jeon S,   Seo J, MacDonald A H,  Bernevig A, and  Yazdani A Science 2014 {\bf 347} 602
\bibitem{morgenstern_2015} Pauly C, Saunus C, Liebmann M, and Morgenstern M 2015 Phys. Rev. B {\bf 92} 085140
\bibitem{yu_2017} Peng  L, Yuan Y, Li G, Yang X, Xian J-J, Yi C-J, Shi Y-G, and Fu Y-S 2017 Nat. Commun. {\bf  8}   659
\bibitem{peregbarnea_2017}   Tiwari K L, Coish W A, and Pereg-Barnea T 2017 Phys. Rev. B {\bf 96}  235120 
\bibitem{yoshino_2019}  Kaku S, Ando T, and  Yoshino J 2019 ACS Nano  {\bf 13} 12980
\bibitem{voigtlander_2021}  Morgenstern M, Pauly C,  Kellner J,  Liebmann M,  Pratzer M, Bihlmayer G,  Eschbach M, Plucinski L, Otto S, Rasche B, Ruck M, Richter M, Just S,  L\"upke F,  and Voigtl\"ander B 2021   Phys. Stat. Sol. {\bf B 258}  2000060 

%%%%%%%%%%%%%%%%%%%%%%%%%%%%%%%%
%%%% quantum computing with topological edge states
%%%%%%%%%%%%%%%%%%%%%%%%%%%%%%
\bibitem{zhang_PNAS}  Liana B,  Sun X-Q,   Vaezi A,   Qi X-L, and   Zhang S-C 2018 Proc. Nat. Acad. Sci. {\bf 115} 10938
\bibitem{he-he_2019} He  M, Sun  H, and He  Q L 2019 Front. Phys. {\bf 14} 43401
\bibitem{guo_2019} Guo X, Hu  G, Zhang  Y, Liu R, Dan  M, Li L, and Zhang  Y 2019 Nano Energy {\bf 60} 36


%%%%%%%%%%%%%%%%%%%%%%%%%%%%%%%%%%%%%%%%%%
%%%%% cold atoms implementations of topological states and phases %%%%%%%
%%%%%%%%%%%%%%%%%%%%%%%%%%%%%%%%%%%%%%%%%%%
\bibitem{dassarma_2008} Scarola  V W, and Das Sarma  S 2008 Phys. Rev. A {\bf 77} 023612
\bibitem{bercioux} Goldman  N,   Urban  D F, and Bercioux  D 2011 Phys. Rev. A {\bf 83} 063601
\bibitem{oh_2011} Mei F, Zhu S-L, Feng X-L, Zhang Z-M, and Oh C H 2011 Phys. Rev. A {\bf  84} 023622
\bibitem{goldman_2012} Mei  F, Zhu S-L, Zhang Z-M, Oh C H, and Goldman N 2012 Phys. Rev. A {\bf 85} 013638
\bibitem{bloch_2013} Atala M, Aidelsburger M, Barreiro  J T,  Abanin D, Kitagawa  T, Demler  E, and Bloch I 2013 Nat. Phys. {\bf 9} 795
\bibitem{lee} Liu X-J,  Law K T, Ng T K,  and Lee P A 2013 Phys. Rev. Lett. {\bf 111} 120402
\bibitem{lecheminant} Nonne  H, Moliner M, Capponi S, Lecheminant P, and Totsuka  K 2013 Eur. Phys. Lett. {\bf 102} 37008
\bibitem{ng}  Liu X-J, Law K T, and Ng T K 2014 Phys. Rev. Lett. {\bf 112} 086401
\bibitem{zoller} Laflamme  C, Baranov  M A, Zoller P, and Kraus  C V 2014 Phys. Rev. A {\bf 89} 022319
\bibitem{deng} Deng  D-L, Wang S-T, and Duan L-M 2014 Phys. Rev. A {\bf 90} 041601
\bibitem{orth} Scheurer M S, Rachel S, and Orth P P 2015 Sci. Rep. {\bf 5} 8386
\bibitem{asboth-lewenstein} Mugel  S, Celi A, Massignan P,  Asb\'oth  J K, Lewenstein M, and Lobo C 2016 Phys. Rev. A {\bf 94} 023631
\bibitem{yang} Zhai H, Rechtsman  M, Lu Y-M, and Yang K 2016 New J. Phys. {\bf 18} 080201
\bibitem{zhu_2017} Mai X-Y, Zhang D-W, Li Z, and Zhu S-L 2017 Phys. Rev. A {\bf 95} 063616
\bibitem{vishwanath} Potirniche  I-D, Potter  A C, Schleier-Smith M, Vishwanath A,  and Yao N Y 2017 Phys. Rev. Lett. {\bf 119} 123601
\bibitem{wu_2017} Ye X-S, Liu Y-J, Zhang X Y, and Wu G 2017 Sci. Rep. {\bf 7} 13541
\bibitem{zhu_2018} Zhang D-W, Zhu Y-Q, Zhao Y X, Yan H, and Zhu S-L, 2018 Adv. Phys. {\bf 67} 253 
\bibitem{an_2019} Liu H, Xiong T-S, Zhang W, and An J-H 2019 Phys. Rev. A {\bf  100} 023622

%%%%%%%%%%%%%%
%%% quenches %%%%%
%%%%%%%%%%%%%%
\bibitem{calabrese_2006} Calabrese  P, and  Cardy J 2006 Phys. Rev. Lett.  {\bf 96}, 136801
\bibitem{polkovnikov_review}  Polkovnikov A,  Sengupta K,  Silva A, and Vengalattore  M  2011 Rev. Mod. Phys.  {\bf 83}, 863.
\bibitem{eisert_2015}  Eisert J, Friesdorf M, Gogolin C  2015 Nat. Phys. {\bf 11} 124.
\bibitem{mitra_2018}  Mitra A 2018  Ann. Rev. Cond. Mat. Phys.  {\bf 9} 245.








%%%%%%%%%%%%%%
%% SSH model
%%%%%%%%%%%%%%
\bibitem{SSH_PRL1979} Su W P, Schrieffer J  R, and  Heeger A J 1979 Phys. Rev. Lett. {\bf 42} 1698
\bibitem{SSH_PRB1980} Su W P, Schrieffer J  R, and  Heeger A J 1980 Phys. Rev. B {\bf 22} 2099
\bibitem{barford_book} Barford W 2005 {\it Electronic and Optical properties of conjugated polymers} (Oxford: Clarendon Press)
%\bibitem{SSH_review}  Heeger A J,   Kivelson S, Schrieffer J  R, Su W P 1988 Rev. Mod. Phys. {\bf 60} 781 

%%%%%%%%%%%%%%%%%%%%%%%%%%%
%%%% SSH model as a topological insulator %%%%
%%%%%%%%%%%%%%%%%%%%%%%%%%%
\bibitem{ungheresi_book}  Asb\'oth J K,    Oroszl\'any L, and  P\'alyi A 2016 {\it A short course on topological Insulators}   (Berlin: Springer)
\bibitem{shen_book} Shen Q-C 2012 {\it Topological Insulators} (Heidelberg: Springer)


%%%%%%%%%%%%%%%%%%%%%%%%%%%
%%% SSH implementations with cold atoms %%%%%
%%%%%%%%%%%%%%%%%%%%%%%%%%%%
\bibitem{gadway_2016}   Meier E J,   An F A, and   Gadway B 2016 Nat.Commun. {\bf 7} 13986
\bibitem{gadway_2019} Xie D, Gou  W, Xiao  T, Gadway  B, and Yan  B 2019 npj Quantum Inform. {\bf 5} 55 

%%%%%%%%%%%%%%%%%%%%%%%%%%%
%%%% time-dependent SSH model %%%%%%%%%
%%%%%%%%%%%%%%%%%%%%%%%%%%%
\bibitem{asboth_PRB_2016}   Boross P,  Asb\'oth J K,  Sz\'echenyi G,  Oroszl\'any L, and P\'alyi A 2016 Phys. Rev. B {\bf 100}  045414  
\bibitem{foatorres_PRA_2015} V. Dal Lago V,  Atala M, and Foa Torres  L E F 2015 Phys. Rev. A {\bf 92} 023624
\bibitem{dutta_2019a}   Bandyopadhyay S, Bhattacharya U, and Dutta A 2019 Phys. Rev. B {\bf 100}  054305
\bibitem{dutta_2019b}   Bandyopadhyay S, and Dutta A 2019 Phys. Rev. B {\bf 100} 144302
\bibitem{xie_2019} L\"u X-L, and  Xie H 2019, J. Phys.: Condens. Matter {\bf 31} 495401




%%%%%%%%%%%%%%%%%%
%%%% Peierls substitution %%%%
%%%%%%%%%%%%%%%%%%
\bibitem{peierls}  Peierls R E  1933 Z. Phys. {\bf 80}, 763 
\bibitem{graf-vogl}  Graf M, and  Vogl P 1995 Phys. Rev. B {\bf 51} 4940


%%%%%%%%%%%%%%%%%%%%%%
%%%% SSH quench in energy space 
%%%%%%%%%%%%%%%%%%%%%%%%
\bibitem{sassetti_PRB2018}   Porta S,  Traverso Ziani N, Kennes D M,  Gambetta F M,  Sassetti M, and Cavaliere F 2018 Phys. Rev. B {\bf 98}, 214306


\bibitem{nota-parity} Specifically, {Eq.}(\ref{SSH-clean}) commutes with $\mathcal{P}$, defined through $\mathcal{P} c_{j,A} \mathcal{P}^{-1}=c_{M-j+1,B}$ and $\mathcal{P} c_{j,B} \mathcal{P}^{-1}=c_{M-j+1,A}$


%%%%%%%%%%%%%%%%%%%
%%% Anderson localization %%%%%
%%%%%%%%%%%%%%%%%%%
\bibitem{anderson_PR_1958} Anderson  P  W  1958 Phys. Rev. {\bf 109} 1492 
\bibitem{thouless_1972}  Thouless  D J   1972  J.   Phys. C: Solid State Phys. {\bf 5}  77




\bibitem{nota_Hchib} The first-quantized version of {Eq.}(\ref{H-chi-b}) is $H_{\chi b}=\oplus_{j=1}^M \delta_j (\sigma_z)_j$ and fulfills $S H_{\chi b} S^{-1}=+H_{\chi b}$, causing the breaking of {Eq.}(\ref{S-def-mat}).
\bibitem{chien_2016}  He Y,  and  Chien C-C 2016 Phys. Rev. B {\bf 94}  024308.




\bibitem{gebhard} Gebhard F, Bott K, Scheidler M, Thomas P, and Koch S W 1997 Phil. Mag. {\bf B75} 1  
\bibitem{TO-GE}   Rossi L,  Dolcini F, Cavaliere F,  Traverso Ziani N,  Sassetti M, and   Rossi F  2021 Entropy {\bf 23}, 220
 
 %%%% dynamical invariants and dynamical quantum phase transitions %%%%%%
 {
\bibitem{Heyl_2013} Heyl M, Polkovnikov A, and Keherein S, 2013 Phys. Rev. Lett. {\bf 110}  135704 
\bibitem{Vajna_2015} Vajna S, and D\'ora B, 2015 Phys. Rev. B {\bf 91} 155127 
 \bibitem{Yang_2018} Yang C, Li L, and Chen S, 2018 Phys. Rev. B {\bf 97}  060304(R)
}

%%%% implementation of an effective flux in optical lattices %%%%%%
\bibitem{spielman_2012}  Jim\'enez-Garc\'ia K,  LeBlanc L J, Williams R A, Beeler M C, Perry A R, and  Spielman I B  2012 Phys. Rev. Lett. {\bf 108}  225303 
\bibitem{lewenstein_2012} Struck J, \"Olschl\"ager C,  Weinberg M,  Hauke P,  Simonet J, Eckardt A,  Lewenstein M, Sengstock K, and  Windpassinger P 2012 Phys. Rev. Lett. {\bf 108} 225304
{\bibitem{paredes_2017} Velasco C G, and Paredes B 2017 Phys. Rev. Lett. {\bf 119} 115301}


%%%%%%%%%%%%%%%%%%%%%%%%
%%%% tenfold classification scheme %%%%%
%%%%%%%%%%%%%%%%%%%%%%%%
\bibitem{nota-su-k-mode-transformation} The  $k$-mode operators $c_{k,s}$ (with $s=A,B=\pm$)  transform under $\mathcal{C},\mathcal{S}$ and $\mathcal{T}$   as follows:  $\mathcal{C} \hat{c}^{}_{k,s}\mathcal{C}^{-1} =(-1)^s \hat{c}^{\dagger}_{-k,s}$,  $\mathcal{S} \hat{c}^{}_{k,s}\mathcal{S}^{-1} =(-1)^s \hat{c}^{\dagger}_{k,s}$ and $\mathcal{T} \hat{c}^{}_{k,s}\mathcal{T}^{-1} = \hat{c}^{}_{-k,s}$.  
\bibitem{ludwig-schnyder-ryu_NJP}    Ryu S,    Schnyder A P,  Furusaki A, and Ludwig A W W 2010 New J. Phys. {\bf 12} 065010
\bibitem{schnyder-ryu_RMP}  Chiu C-K, J C Y Teo, Ryu S,    Schnyder A P  2016 Rev. Mod. Phys. {\bf 88} 035005

%%%%%%%%%%%%%%%%%%%%%%%%
%% exact solutions for the ring and the chain
%%%%%%%%%%%%%%%%%%%%%%%
\bibitem{montambaux_2011} Delplace P, Ullmo D, and Montambaux G 2011  Phys. Rev. B {\bf 84}  195452
\bibitem{hilke-mackenzie_pla_2021}  Zaimi M, Boudreault Ch,  Baspin N,  Delnour N,  Eleuch H,
MacKenzie R, and  Hilke M 2021  Phys. Lett. {\bf A388} 127035





\end{thebibliography}
\end{document}